\documentclass[a4paper,11pt]{article}
\usepackage{jheppub}
\usepackage{lineno}

\usepackage{graphicx} %
\usepackage{hyperref}
\usepackage{amsmath}
\usepackage{amsfonts}
\usepackage{amssymb}
\usepackage{slashed}
\usepackage{physics}
\usepackage{float}
\usepackage{caption, subcaption}
\usepackage{booktabs}

\usepackage{tikz}
\usetikzlibrary {arrows.meta,bending,positioning}
\usetikzlibrary{angles}

\title{Underground Production of Electromagnetic Dark States by MeV-scale Electron Beams and Detection with CCDs}

\author[a]{Helmut Eberl,}
\author[a,b,c]{Maximilian Fahrecker,}
\author[a,b]{Josef Pradler}

\affiliation[a]{
Marietta Blau Institute for Particle Physics, Austrian Academy of Sciences, Dominikanerbastei 16, A-1010 Vienna, Austria}
\affiliation[b]{University of Vienna, Faculty of Physics, Boltzmanngasse 5, A-1090 Vienna, Austria}
\affiliation[c]{University of Vienna, Vienna Doctoral School in Physics, Boltzmanngasse 5, 1090 Vienna, Austria}
\emailAdd{maximilian.fahrecker@univie.ac.at}

\abstract{In this work we explore the possibility of new light fermionic particles with millicharge or electromagnetic form factor interactions and their underground production via an electron beam in the 100\,MeV range and their subsequent detection using a CCD-sensor. We evaluate the S-matrix elements and the phase spaces for production analytically, and then calculate the corresponding cross sections numerically. For millicharged fermions this set-up could be able to probe a window in parameter space, yet unconstrained by direct detection experiments. The electric or magnetic dipole moment of a light fermion could feasibly be probed with enough beam time or an increased beam energy.}

\begin{document}

\maketitle

\section{Introduction}

The nature of cold dark matter (DM), required by the cosmological standard model $\Lambda$CDM~\cite{WMAP:2012nax,Planck:2018vyg}, remains one of the most pressing questions in modern physics. While there is solid indirect evidence for DM from cosmic microwave background measurements and astrophysical observations, no conclusive direct detection of a DM particle has yet been achieved. DM candidates are expected to be particles beyond the Standard Model (SM). Although a multitude of extensions to the SM have been proposed, many are already tightly constrained by existing measurements. For example, weakly interacting massive particles (WIMPs) and other heavy DM candidates, such as supersymmetric particles, have been intensely studied over the past decades~\cite{Bertone:2004pz,Bertone:2010zza,XENON:2018voc,Billard:2021uyg}. Light, non-axion, sub-GeV dark matter has attracted particular interest in recent years due to its relatively unexplored parameter space. A number of direct searches targeting this mass range have been conducted or proposed~\cite{Essig:2011nj,CRESST:2019jnq,EDELWEISS:2020fxc,LZ:2023poo}. Of particular interest are low-threshold semiconductor searches employing CCD detectors~\cite{Chavarria:2014ika,Essig:2015cda,DAMIC:2016lrs,Tiffenberg:2017aac,Crisler:2018gci,DAMIC:2019dcn,SENSEI:2019ibb,SENSEI:2020dpa,DAMIC:2020cut,Castello-Mor:2020jhd,CONNIE:2019xid,Oscura:2022vmi,SENSEI:2023gie,DAMIC-M:2023gxo,Essig:2024dpa}.

A major challenge in the direct detection of light dark matter arises from its low kinetic energy in the Galactic halo compared to WIMPs. For instance, in our local region of the halo, a DM particle with mass $1\:\mathrm{MeV}$ possesses only a few~eV of kinetic energy, given a typical velocity of about $230\:\mathrm{km/s}$~\cite{deSalas:2020hbh}. This limitation can be addressed either by lowering experimental energy thresholds or by producing DM particles with higher velocities. For DM gravitationally bound to the Galaxy, only the first approach is feasible, requiring detectors with extreme sensitivity to small energy deposits. Alternatively, one can produce DM particles with large kinetic energies, rendering even sub-MeV masses detectable through their relativistic interactions. Accelerator-based experiments are particularly well suited to this task, and several have been conducted~\cite{Prinz:1998ua,CMS:2012xi,Marocco:2020dqu} or proposed~\cite{Kahn:2014sra,BDX:2016akw,LDMX:2018cma,Arguelles:2019xgp,Kim:2021eix,milliQan:2021lne,Essig:2024dpa}.

Dark matter particles can be produced using an accelerator or collider. In this work, we focus on a linear electron accelerator in which the beam is directed onto a dump. The incident electrons interact with nuclei in the dump, producing DM particles according to the underlying model. A detector positioned downstream along the beam line can then search for the resulting, ideally relativistic, DM flux. To ensure sensitivity to dark states, backgrounds from standard particles must be strongly suppressed. This is achieved by placing appropriate shielding in the beam line so that primarily weakly interacting particles, such as DM, reach the detector, while other radiation is minimized. The entire accelerator-detector set-up should be located in an underground laboratory, as is typical for dark matter searches. In the present study, we consider beam energies of order $100\,\mathrm{MeV}$, comparatively low for a fixed-target experiment~\cite{Prinz:1998ua,LDMX:2018cma}, which makes such a set-up potentially compatible with existing underground detection facilities, provided they can accommodate a compact electron accelerator. This work is intended as a conceptual, theory-driven exploration of such a configuration, with the aim of informing experimental studies of sub-GeV electron beam-dump experiments employing CCD detector technology.

In summary, our calculation of dark state production focuses on the energy and emission angle of the produced particles relative to the beam line, which determine the incident energy on the detector. For detection, the key observable is the recoil energy of the target material, corresponding directly to the energy deposited by the DM. While we frame our discussion in terms of dark matter, the same framework applies more generally to the production of other dark-sector states, including unstable particles with sufficiently long lifetimes to traverse the shielding and reach the detector. The essential kinematic quantities governing the full production–detection process are the electron and DM masses, together with the beam energy, set by the electron momentum. Using these parameters, we compute the expected event rate in the detector. The geometry of our set-up is similar to that of the SLAC-mQ experiment~\cite{Prinz:1998ua}, though employing a less energetic beam and a different detector design. While~\cite{Essig:2024dpa} explores a comparable configuration, our study uses beam energies lower by three orders of magnitude and performs all phase-space calculations analytically rather than via event generators.

\section{Particle Production by Electron Beams}

\subsection{Electromagnetic Form Factor Interactions}
In this work we will specifically investigate a new light fermionic particle $\chi$ with mass $m_{\chi}$ and proposed millicharge or electromagnetic (EM) form factor couplings~\cite{Pospelov:2000bq,Sigurdson:2004zp,Barger:2010gv,Ho:2012bg}. The term millicharge refers to a fractional electric charge that can arise from dark photon models~\cite{Holdom:1985ag,Goldberg:1986nk,Fabbrichesi:2020wbt} or appear in (supersymmetric) grand unified theories~\cite{Fayet:1980ad,Okun:1983vw}. Electromagnetic form factors are the low energy Wilson coefficients of an effective field theory. The phenomenology and viability of particles with millicharge~\cite{Golowich:1986tj,Prinz:1998ua, Davidson:2000hf, Dubovsky:2003yn, McDermott:2010pa, Cline:2012is, Dolgov:2013una, Izaguirre:2013uxa, Vogel:2013raa, Dvorkin:2013cea, Soper:2014ska, Ali-Haimoud:2015pwa, Kamada:2016qjo, Munoz:2018pzp, Berlin:2018sjs, Barkana:2018qrx, Chang:2018rso, Kovetz:2018zan, Xu:2018efh, Slatyer:2018aqg, Berlin:2018bsc, Magill:2018tbb, An:2021qdl, Buen-Abad:2021mvc, Chu:2023jyb, Gan:2023jbs} or electromagnetic form factor interactions~\cite{Pospelov:2005pr, Schmidt:2012yg, Kopp:2014tsa, Ibarra:2015fqa, Kavanagh:2018xeh,Chu:2018qrm,Chu:2019rok,Trickle:2019ovy,Chu:2020ysb, Hambye:2021xvd, PandaX:2023toi,Chu:2023zbo} has already been explored. We will explore the parameter space independent of the relic density constraint. The relevant Lagrangians interaction terms for a Dirac fermion~$\chi$ read~\cite{Chu:2018qrm}:
\begin{subequations}
	\label{eq:mod.1}
	\begin{align}
	\label{eq:mod.1a} %
	&\text{millicharge $(\varepsilon e)$:} ~~~~ &\mathcal{L}_{\varepsilon e} &= \varepsilon e \: \bar{\chi} \gamma^{\mu} \chi A_{\mu},
	\\
	\label{eq:mod.1b} %
	&\text{magnetic dipole moment - MDM $(\mu_{\chi})$:} ~~~~ &\mathcal{L}_{\text{MDM}} &= \frac{1}{2} \mu_{\chi} \: \bar{\chi} \sigma^{\mu\nu} \chi F_{\mu\nu},
	\\
	\label{eq:mod.1c} %
	&\text{electric dipole moment - EDM $(d_{\chi})$:} ~~~~ &\mathcal{L}_{\text{EDM}} &= \frac{i}{2} d_{\chi} \: \bar{\chi} \sigma^{\mu\nu} \gamma^5 \chi F_{\mu\nu},
	\\
	\label{eq:mod.1d} %
	&\text{anapole moment - AM $(a_{\chi})$:} ~~~~ &\mathcal{L}_{\text{AM}} &= \vphantom{\frac{1}{2}} - a_{\chi} \: \bar{\chi} \gamma^{\mu} \gamma^5 \chi \partial^{\nu} F_{\mu\nu},
	\\
	\label{eq:mod.1e} %
	&\text{charge radius - CR $(b_{\chi})$:} ~~~~ &\mathcal{L}_{\text{CR}} &= b_{\chi} \: \bar{\chi} \gamma^{\mu} \chi \partial^{\nu} F_{\mu\nu},
	\end{align}
\end{subequations}
where $F_{\mu\nu}=\partial_{\mu}A_\nu-\partial_{\nu}A_\mu$ is the electromagnetic field strength tensor, $\sigma^{\mu\nu}=\frac{i}{2}\comm{\gamma^{\mu}}{\gamma^{\nu}}$ and  ${\gamma^5=i\gamma^0\gamma^1\gamma^2\gamma^3}$. 
For the mass-dimension-5 operators, $\mu_{\chi}$ is the magnetic dipole moment (MDM) and $d_{\chi}$ the electric dipole moment (EDM). The mass-dimension-6 operator couplings  $a_{\chi}$ and $b_{\chi}$ are called anapole moment (AM)~\cite{1958JETP....6.1184Z} and charge radius (CR), respectively. These couplings are all real-valued. Since the AM and CR operators contain the derivative of the photon field strength, they vanish for interactions with on-shell photons. There is a variety of UV-complete theories that can generate these form factors. For instance, EM moments can arise from loops of charged particles~\cite{Raby:1987ga, Pospelov:2008qx}. A simple example of this is a model with an additional scalar, charged under the SM U(1) hypercharge field, that can induce EM moments through loops~\cite{Barger:2010gv,Schmidt:2012yg, Kopp:2014tsa, Ibarra:2015fqa,Kavanagh:2018xeh}. In addition, the particle $\chi$ need not be fundamental and can have form factors due to its internal structure, like the neutron~\cite{Bagnasco:1993st,Foadi:2008qv,Antipin:2015xia}.

At mass-dimension-7 so-called Rayleigh or susceptibility interactions enter, which are proportional to the square of the (dual of the) EM field strength tensor~\cite{Weiner:2012cb}. These will not be considered here, since they have an additional suppression by the effective energy scale. Furthermore, these interactions would only contribute to exclusive $\chi$ production at 1-loop order, leading to further suppression.

\subsection{Electron-Electron Bremsstrahlung Pair Production}

\begin{figure}%
	\centering
	\begin{subfigure}[h]{0.45\textwidth}
        \centering
		\includegraphics[width=0.67\textwidth, height=!]{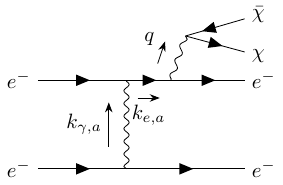}
        \caption{}
	\end{subfigure}
	\begin{subfigure}[h]{0.45\textwidth}
		\centering
		\includegraphics[width=0.67\textwidth, height=!]{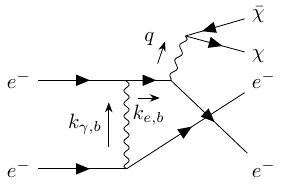}
        \caption{}
	\end{subfigure} %
	\begin{subfigure}[h]{0.45\textwidth}
		\centering
		\includegraphics[width=0.67\textwidth, height=!]{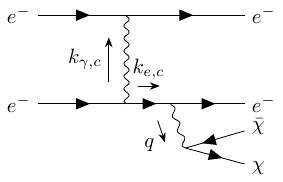}
        \caption{}
	\end{subfigure} %
	\begin{subfigure}[h]{0.45\textwidth}
		\centering
		\includegraphics[width=0.67\textwidth, height=!]{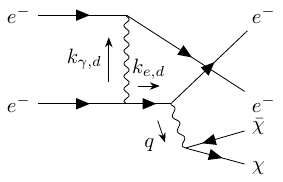}
        \caption{}
	\end{subfigure} %
	\caption{Production of $\chi\bar{\chi}$ pairs from a photon emitted by an outgoing electron. The $k_{\gamma,n}$ and $k_{e,n}$, with ${n\in\{a,b,c,d\}}$, represent the momenta for each diagram indexed by $n$.}
	\label{fig:prodFeyn.f} %
\end{figure}

The main mode of production is bremsstrahlung from electron-electron scattering. %
At higher energies, meson decays play an important role in dark state production at electron beams~\cite{Essig:2024dpa} as well. 
With a beam energy of 100\,MeV, we remain well below the di-meson production threshold. The main focus of this work will hence be the production through electron-electron bremsstrahlung.
The associated $\chi$-energy and $\chi$-angle differential production cross section in the lab frame, i.e. the rest frame of the beam dump electrons, is given by
\begin{equation}
\frac{d^2\sigma_{\text{prod}}}{d E_{\chi} d\cos\theta_{\chi}} = \frac{\abs{\vec{p}_{\chi}}}{4m_e\abs{\vec{p}_2}}  \int \frac{d\Pi_{2\rightarrow 4}}{ds_{34\bar{\chi}} dt_{2\chi}} \frac{1}{\abs{J}} \abs{\mathcal{M}_{2\rightarrow 4}}^2 , 
\label{eq:pcs.1} %
\end{equation}
where $\abs{\mathcal{M}_{2\rightarrow 4}}^2$ is the fully spin-summed squared matrix element of the 2-to-4 particle production process, $1/(4m_e\abs{\vec{p}_2})$ is the flux factor in the lab frame, with $\vec{p}_2$ denoting the momentum of the incoming beam electron with an energy of $E_2$. The factor $\abs{\vec{p}_{\chi}}/\abs{J}$ comes from the transformation of the invariants $(s_{34\bar{\chi}}, t_{2\chi})$ to the lab frame variables $(E_{\chi}, \cos\theta_{\chi})$; the Jacobian $\abs{J}$ will cancel with the one inside the phase space integral~(\ref{eq:fphi.7}). For a detailed analytical derivation of the phase space, see App.~\ref{app:b}.

The relevant Feynman diagrams for the pair production of $\chi$ using electron-electron scattering are shown in Fig.~\ref{fig:prodFeyn.f} for a photon radiated from an outgoing electron, and in Fig.~\ref{fig:prodFeyn.i} for a photon radiated by an incoming electron. They correspond to bremsstrahlung emission of a virtual photon, which subsequently produces a $\chi$ particle-antiparticle pair. The momentum transfer $k_{\gamma}$ and the electron propagator momentum $k_{e}$ in these figures depend on the particular diagram, while $q=p_{\chi}+p_{\bar{\chi}}$ is the total momentum of the $\chi$ particle-antiparticle pair. Accounting for exchange diagrams, there is a total of eight Feynman diagrams to consider, since the scattering partners are identical, see App.~\ref{app:c}. This also prevents us from neglecting the projectile mass, as is usually done in the context of electron-nucleus scattering.%
\footnote[1]{\label{fn:1}It can be seen from kinematic considerations, (\ref{eq:intlim.mchimax}) and following, that in case of equal-mass fixed-target scattering, setting $m_e=0$ is forbidden since $\lim_{m_e\to0} E_{\chi}^{\rm min/max}$ becomes indeterminate. In other words, the lab frame, i.e. the frame in which on electron is at rest, does not exist for $m_e=0$. Therefore, while it is not strictly forbidden to set $m_e=0$, one would have to reintroduce the mass after calculating the cross section in the centre-of-mass frame, since it is necessary to boost into the lab frame to determine the angular region for the detector. Additionally, in the case where $m_{\chi}\sim m_e$ or $m_{\chi}<m_e$, neglecting $m_e$ would introduce large (numeric) errors as $m_e$ gives the dominant mass scale or is at least of similar magnitude as~$m_{\chi}$.} 
Additional tree level diagrams arise through a dark current inserted into the $t$- and $u$-channel photon propagators. Such diagrams are, however, of higher order in the EM form factor couplings and can therefore be neglected.

\begin{figure}%
	\centering
	\begin{subfigure}[h]{0.495\textwidth}
		\centering
		\includegraphics[width=0.99\textwidth, height=!]{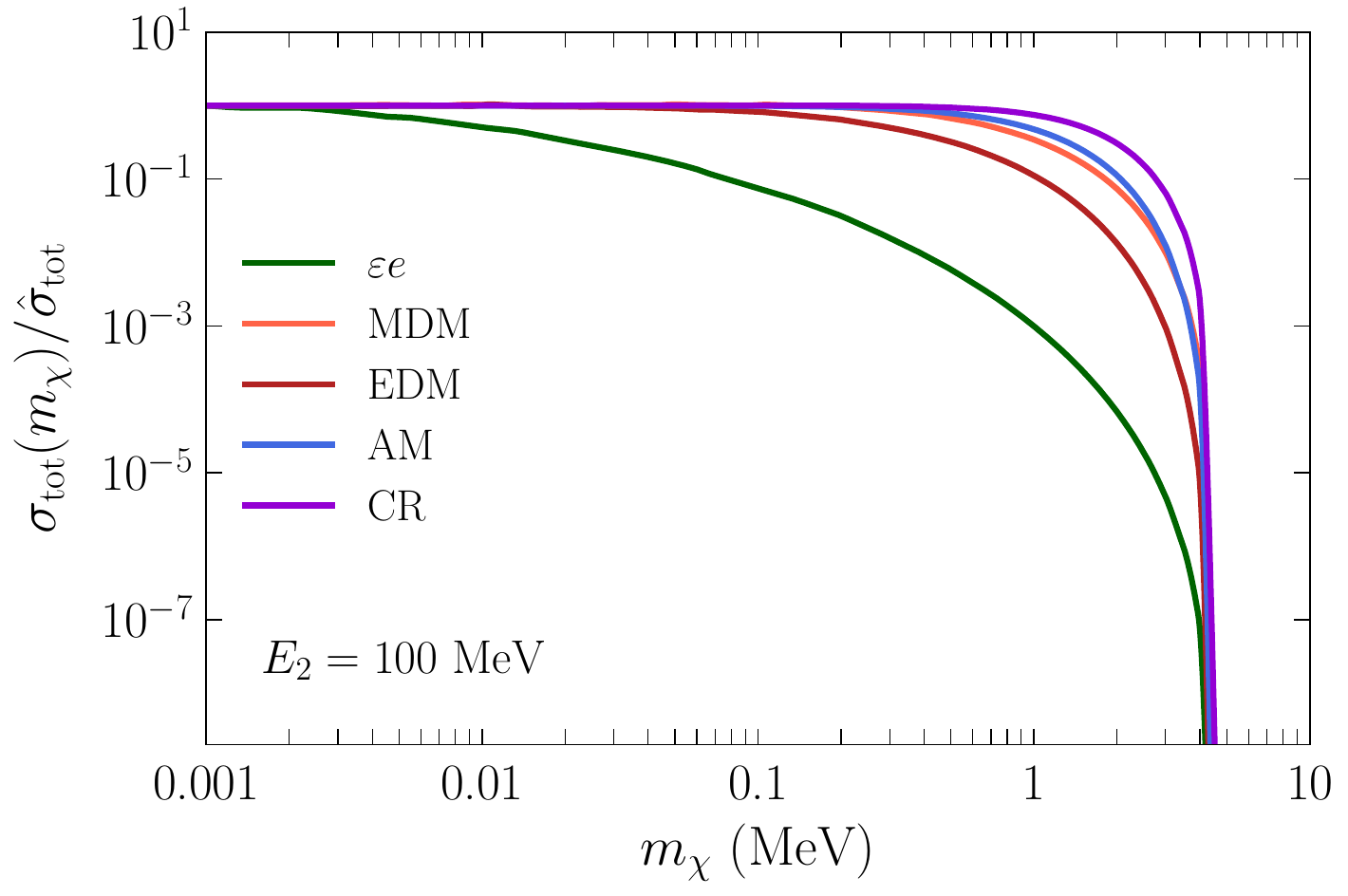}
	\end{subfigure}
	\begin{subfigure}[h]{0.495\textwidth}
		\centering
		\includegraphics[width=0.99\textwidth, height=!]{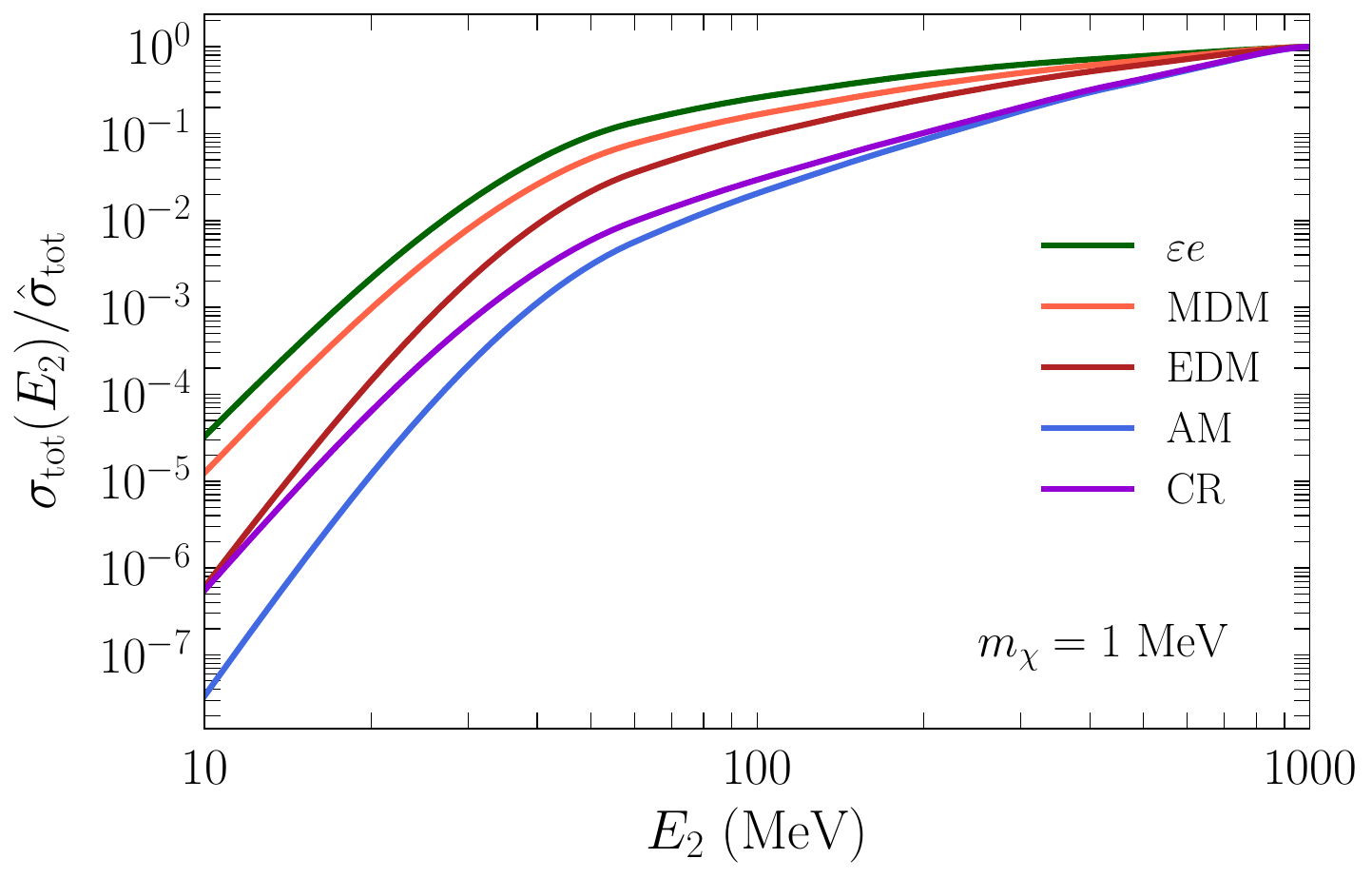}
	\end{subfigure}
	\caption{Total cross section for $\chi\bar{\chi}$ pair production from $e^- e^-$ scattering for all interactions considered, relative to the a reference cross section $\hat{\sigma}_{\text{tot}}$ of the same interaction type. The \emph{left panel} shows the mass dependence of the production cross section at an energy of the incoming electron of $E_2 = 100$\,MeV relative to a reference value $\hat{\sigma}_{\text{tot}}(m_{\chi,\text{ref}})$ at $m_{\chi,\text{ref}}=1$\,keV. The weakening sensitivity to mass with the increasing order of the operators can be seen. The \emph{right panel} shows the dependence of the production cross section on the energy of the incoming electron $E_2$ of a particle with mass $m_{\chi}=1$\,MeV relative to a reference value $\hat{\sigma}_{\text{tot}}(E_{2,\text{ref}})$ at $E_{2,\rm ref}=1$\,GeV. It shows that with increasing operator dimension, the energy scaling becomes more pronounced. Additionally, the operators containing $\gamma^5$ display a steeper fall-off at low input energies. See Tab.~\ref{tab:ref} for the values of the reference cross sections. }
	\label{fig:prodcross}
\end{figure}

The behaviour of total production cross section $\sigma_{\text{prod}}$~(\ref{eq:pcs.1}) for the form factor interactions~(\ref{eq:mod.3b}-\ref{eq:mod.3e}), as a function of the mass of $\chi$ or incoming electron energy $E_2$, are given in Fig.~\ref{fig:prodcross}.
The cross section for millicharge behaves like electron-positron pair production. The effective operators show the expected loss of mass dependence at low $m_{\chi}$. MDM and EDM converge for small mass, confirming that these two interactions only differ by mass-dependent terms. Likewise, for AM and CR this is even more pronounced. The distinct behaviour of two operators related by a $\gamma^5$ insertion arises from their different velocity dependence in the non-relativistic limit. Consequently, near the kinematic endpoint, where the velocity of the produced particles decreases significantly, this effect becomes more pronounced, leading to the noticeably different scaling of interactions containing a $\gamma^5$ at lower energies or higher masses in Fig.~\ref{fig:prodcross}.

\section{Detection by Elastic Scattering in CCDs}

\begin{figure}%
	\centering
	\begin{subfigure}[h]{0.495\textwidth}
		\centering
		\includegraphics[width=0.99\textwidth, height=!]{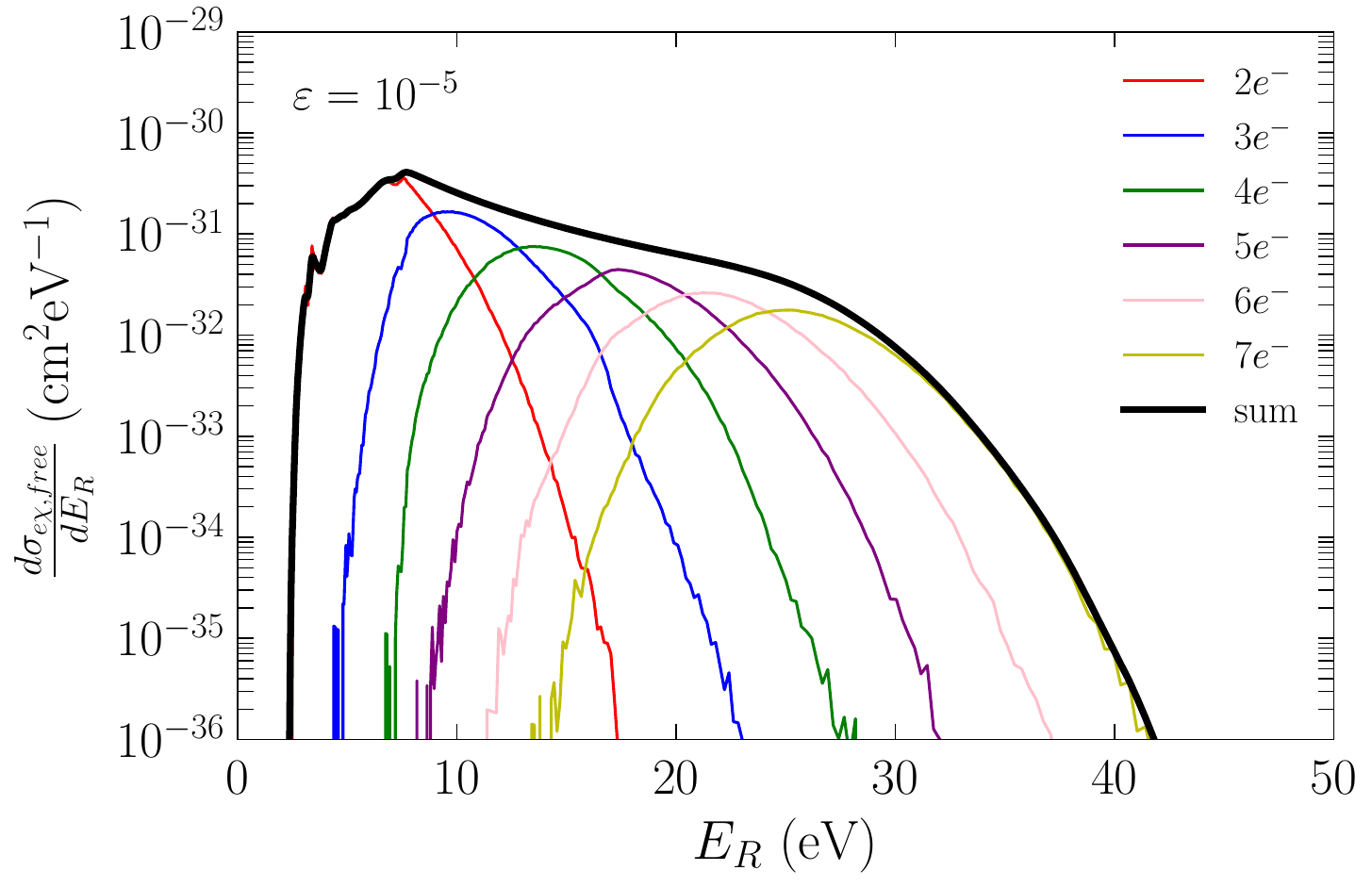}
	\end{subfigure}
	\begin{subfigure}[h]{0.495\textwidth}
		\centering
		\includegraphics[width=0.99\textwidth, height=!]{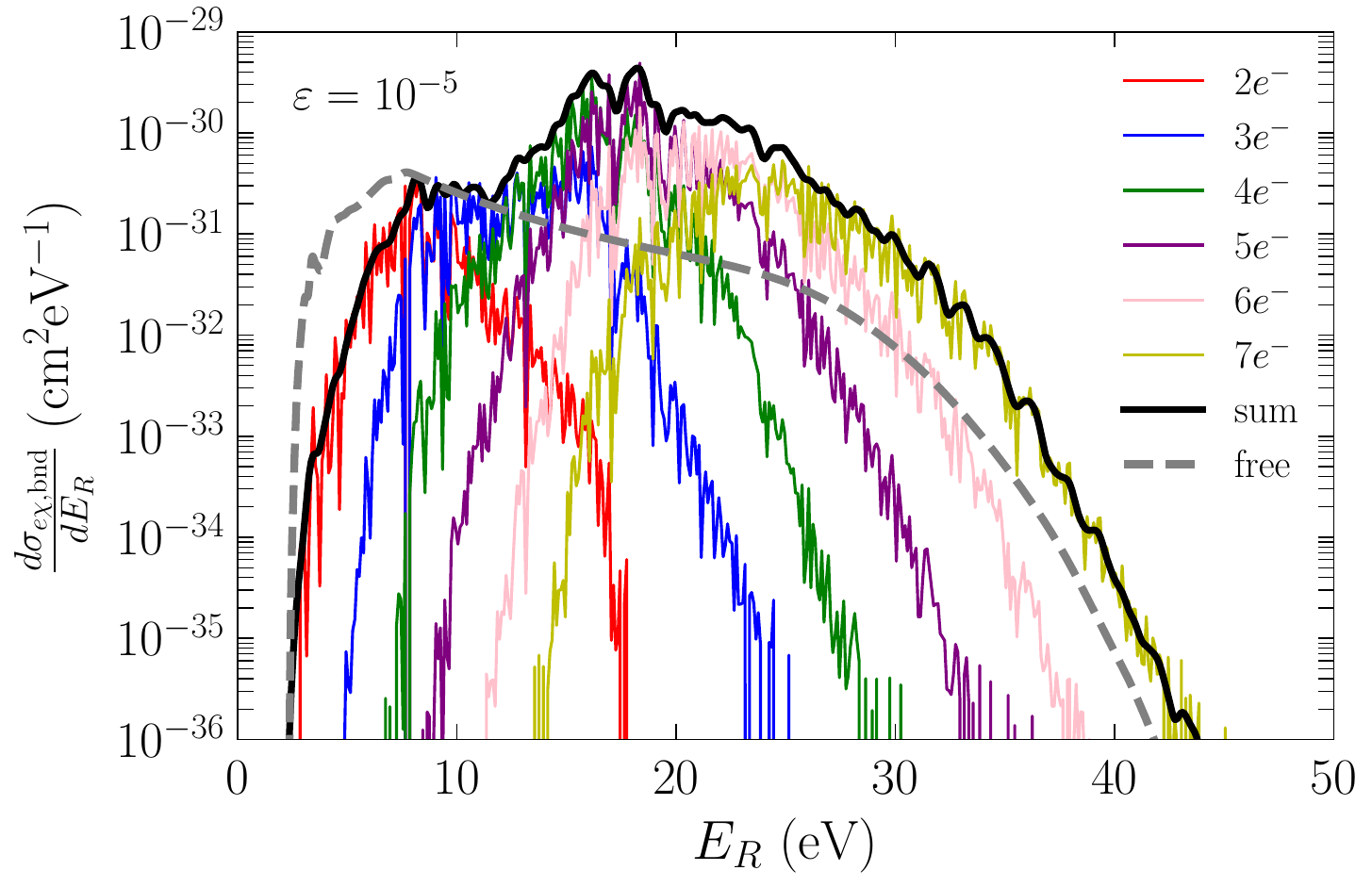}
	\end{subfigure}
	\caption{Differential cross section for the creation of two to seven charges with respect to the recoil energy deposited into electrons by millicharged $\chi$ of energy $E_{\chi}=5$\,MeV and mass $m_{\chi}=1$\,keV. The \emph{left panel} shows the values for free electron-$\chi$ scattering and its sum given by the black solid line. The \emph{right panel} shows the results using an ab-initio electron-loss function of silicon~\cite{Dreyer:2023ovn}. Here, the black line is the (Gaussian-smoothed) sum of bins. Noticeable is the peak at the plasmon resonance frequency for 4 to 5 electron events and the overall larger cross section on the right side compared to the simpler, free-electron case; the latter is shown for better comparison by the dashed grey line.}
	\label{fig:binned_comp}
\end{figure}

The particles produced in the beam dump subsequently travel to the CCD detector where they will liberate charges in its bulk. A straightforward way to estimate this interaction is using free electron-$\chi$ scattering as described in App.~\ref{app:a}. This is shown in the left panel of Fig.~\ref{fig:binned_comp}. However, solid state effects, in particular plasmon interactions, can enhance the detection cross section for high energy particles. One can account for these effects using a dielectric function $\epsilon(\omega,k)$ that models the collective behaviour of the electrons in the detector. Following~\cite{Fermi:1940zz,Allison:1980vw,Essig:2024ebk}, it is possible to write the differential cross section of $\chi$-particles scattering with bound electrons as
\begin{align}
\frac{d\sigma_{e\chi,\rm bnd}}{dE_{R}} = \frac{e^2}{2n\pi^2\beta^2}\int_{E_{R}/\beta}^{2\abs{\vec{p}_{\chi}}-E_{R}/\beta} \frac{dk}{k}& F(E_{R},k) \bigg\{ \text{Im}\left(\frac{-1}{\epsilon(E_{R},k)}\right) \nonumber \\
&+ (\beta^2k^2 - E_{R}^2)\text{Im}\left(\frac{-1}{k^2-E_{R}^2\epsilon(E_{R},k)}\right) \bigg\},
\label{eq:pdet.ss} 
\end{align}
where $E_{R}$ refers to the electron recoil energy, $k$ is the momentum transfer, $\beta=\abs{\vec{p}_{\chi}}/E_\chi$ is the velocity of the incoming $\chi$ particle and $n_e$ is the electron density of the detector material. The crucial part of~(\ref{eq:pdet.ss}) is the inclusion of the imaginary part of the inverse dielectric function $\text{Im}(-1/\epsilon(E_{R},k)) = \epsilon_2(E_{R},k) / |\epsilon(E_{R},k)|^2$, the electron loss function. While this term describes direct Coulomb interactions between electrons, the term on the second line of~(\ref{eq:pdet.ss}) accounts for secondary, indirect energy exchange via photons. The imaginary part of the dielectric function $\epsilon_2(E_{R},k)$ is calculated via the crystal form factor of silicon, which is a DM-independent quantity that encodes the electronic structure of a material and can be computed via lattice~\cite{Essig:2015cda}; in this work  we use \texttt{QCDark}~\cite{Dreyer:2023ovn} for obtaining  $\epsilon_2(E_{R},k)$. This needs to be combined with a method for evaluating the absolute value of the dielectric function $|\epsilon(E_{R},k)|^2$, which we obtain from the Lindhard model (\ref{eq:lindhard}). The different form factor interactions are accounted for by $F(E_{R},k)$~\cite{Essig:2024ebk} in~(\ref{eq:pdet.ss}). These are derived by comparing the matrix elements of Coulomb scattering with those of the EM form factors and are
\begin{subequations}
\begin{alignat}{2}
	\label{eq:pdet.ffa}
	F_{\varepsilon e} &= \varepsilon^2,
	\\
    \label{eq:pdet.ffb}
    F_{\text{MDM}}/\mu_{\chi}^2 = F_{\text{EDM}}/d_{\chi}^2 &= (k^2 -E_{R}^2) /e^2,
	\\
    \label{eq:pdet.ffd}
     F_{\text{AM}}/a_{\chi}^2 = F_{\text{CR}}/b_{\chi}^2 &= (k^2 -E_{R}^2)^2 /e^2,
\end{alignat}
\end{subequations}

When using the $e^-\chi$ scattering cross section given in (\ref{eq:echiES.10}) or (\ref{eq:pdet.ss}), we obtain the rate for energy deposition into an individual electron. Since CCD detectors are able to achieve single electron resolution, it is necessary to model the signal yield for multiple electron events resulting from this energy deposition. For this, the results from~\cite{Ramanathan:2020fwm} are used. In that work, the authors employ a phenomenological description of electron impact ionisation in the important energy range of about $12$\,eV$-50$\,eV that was largely unaccounted for by direct measurements. This is done in order to model the response to energy deposition into silicon and describe how many charges are liberated. The probability for $n$ electron-hole pairs to form when an energy $E_R$ is deposited into the silicon bulk of a CCD is represented by a piecewise function $p_n(E_R)$ that takes precomputed values for a material temperature of 100\,K, taken from a table attached to~\cite{Ramanathan:2020fwm}, for energies $1.1$\,eV$-50$\,eV and follows a Gaussian  model for higher energies. Accordingly, the cross section for producing an $n$-electron event is given by
\begin{equation}
\sigma_{\text{det},n} = \int_{E_R^{\text{min}}}^{E_R^{\text{max}}} dE_R\,p_n(E_R) \frac{d\sigma_{e\chi,\text{bnd}}}{dE_R},
\label{eq:pdet.bin} 
\end{equation}
where $d\sigma_{e\chi,\text{bnd}}/dE_R$ is given in~(\ref{eq:pdet.ss}); for evaluating the detection cross section from the scattering on free electrons, one uses (\ref{eq:echiES.10}) instead. The minimum energy deposition contributing in~(\ref{eq:pdet.bin}) is $E_R^{\text{min}}=1.1\,$eV, the band gap of silicon~\cite{ParticleDataGroup:2024cfk}, and the maximum energy $E_R^{\text{max}}$ is determined by kinematics. In case of free electron scattering it is~(\ref{eq:echiES.11}), while for bound electrons it is determined by an effective cut-off at which the cross section becomes negligible, here $E_R^{\text{max}}=50$\,eV. As can be seen in Fig.~\ref{fig:binned_comp} for millicharge, or in Fig.~\ref{fig:binned_comp_fm} for MDM and AM, bound electron scattering has a larger cross section for the charge event range where Skipper-CCD is most sensitive, mainly due to low background~\cite{SENSEI:2023gie,DAMIC-M:2023gxo}. Therefore, we will focus on the bound state case instead of free scattering in the remainder of this work.

The objective is to detect the $\chi$ particles that are produced by the electron beam. We assume the beam dump is a thick target with respect to its radiation length $X_T$. The energy-differential total number of $n$-charge signal events after $N_\text{EOT}$ electrons on target is then given by
\begin{equation}
\frac{d N_{\text{sig},n} }{dE_{R}} = N_\text{EOT} n_T X_T n_{\text{det}} L_{\text{det}} \int dE_{\chi} d\cos\theta_{\chi} \frac{d^2 \sigma_{\text{prod}}}{dE_{\chi} d\cos\theta_{\chi}} \frac{d\sigma_{\text{det},n}}{dE_R} .
\label{eq:pdet.6} 
\end{equation}
Here, $n_{T}$ is the electron density of the target, $n_{\rm det}$ is the electron density and $L_{\rm det}$ is the length of the detector, respectively. To obtain the number of liberated charges in a bin $N_{\text{bin,charge}}$, one multiplies the event count with the number of electrons $N_{\text{n,charge}}=n N_{\text{sig},n}$. Depending on the detector and the read-out method, it is possible to observe the events in individual charge bins or the total events
\begin{equation}
N_{\text{sig}}  = \sum_{n=2}^{7} N_{\text{sig},n}  ,
\label{eq:pdet.tot} 
\end{equation}
where we have chosen the values for $n$ such that single electron events, which are difficult to distinguish from thermal noise, and higher charge events that are unlikely to be caused by DM, are excluded. Specifically, the $1e^-$ bin is excluded because it is dominated by dark-current background in CCD detectors. For DAMIC-M, for instance, the region of least background is $4e^-$ to $7e^-$ events~\cite{DAMIC-M:2023gxo}, lining up well with the plasmon peak on the right side of Fig~\ref{fig:binned_comp} and motivating the choice of bins in~\eqref{eq:pdet.tot}.

\section{Results}

With the results of~(\ref{eq:pdet.6}), one can compute the total amount of signal events $N_{\text{sig}}$ by integrating it with respect to $\cos\theta_{\chi}$ and $E_R$. In order to obtain predictions for an experimental set-up of production and detection, it is necessary to use some physical parameters. For the beam dump, lead is used as an exemplary material with a radiation length $X_T=0.56$\,cm. Furthermore, an electron beam is needed for the set-up. Here we have used $N_{\text{EOT}}=10^{20}$ as a reference for the total electrons on target, which could be achieved by a 10\,kW beam accelerating electrons with 100\,MV for a total duration of approximately~45\,h.

\begin{figure}%
	\centering
	\includegraphics[width=0.80\textwidth, height=!]{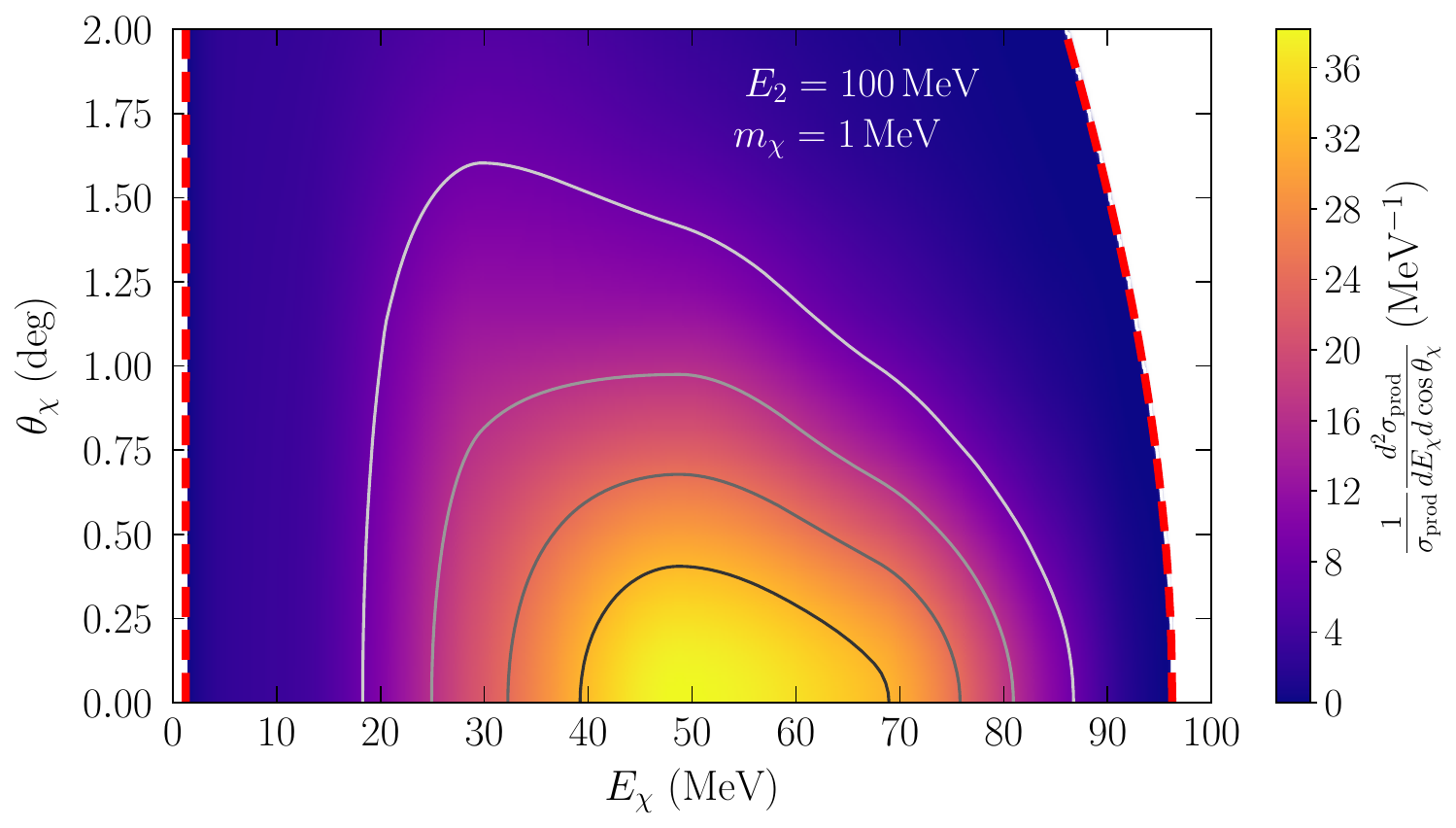}
	\caption{Heat map of the normalised production cross section $\sigma_{\rm prod}^{-1} d^2\sigma_{\rm prod}/dE_\chi d\cos\theta_\chi$ in units of $\rm MeV^{-1}$ shown as a function of the final state $\chi$-energy $E_\chi$ and  production angle $\theta_\chi$ for a millicharged particle of mass $m_\chi = 1~{\rm MeV}$; the beam energy is $E_2=100$\,MeV. Note that the y-scale is displayed in degrees for ease of readability and that the distribution is symmetric for negative values~$\theta<0$. The dominant angular region is between $0^\circ$ to $2^\circ$ and peaks at energies of around 50\,MeV. This is expected for highly relativistic processes, since the particles are boosted into the beam line. The thick red-dashed lines indicate the kinematically forbidden regions.}
	\label{fig:milli.geo}
\end{figure}

\begin{figure}
	\centering
	\includegraphics[width=0.65\textwidth, height=!]{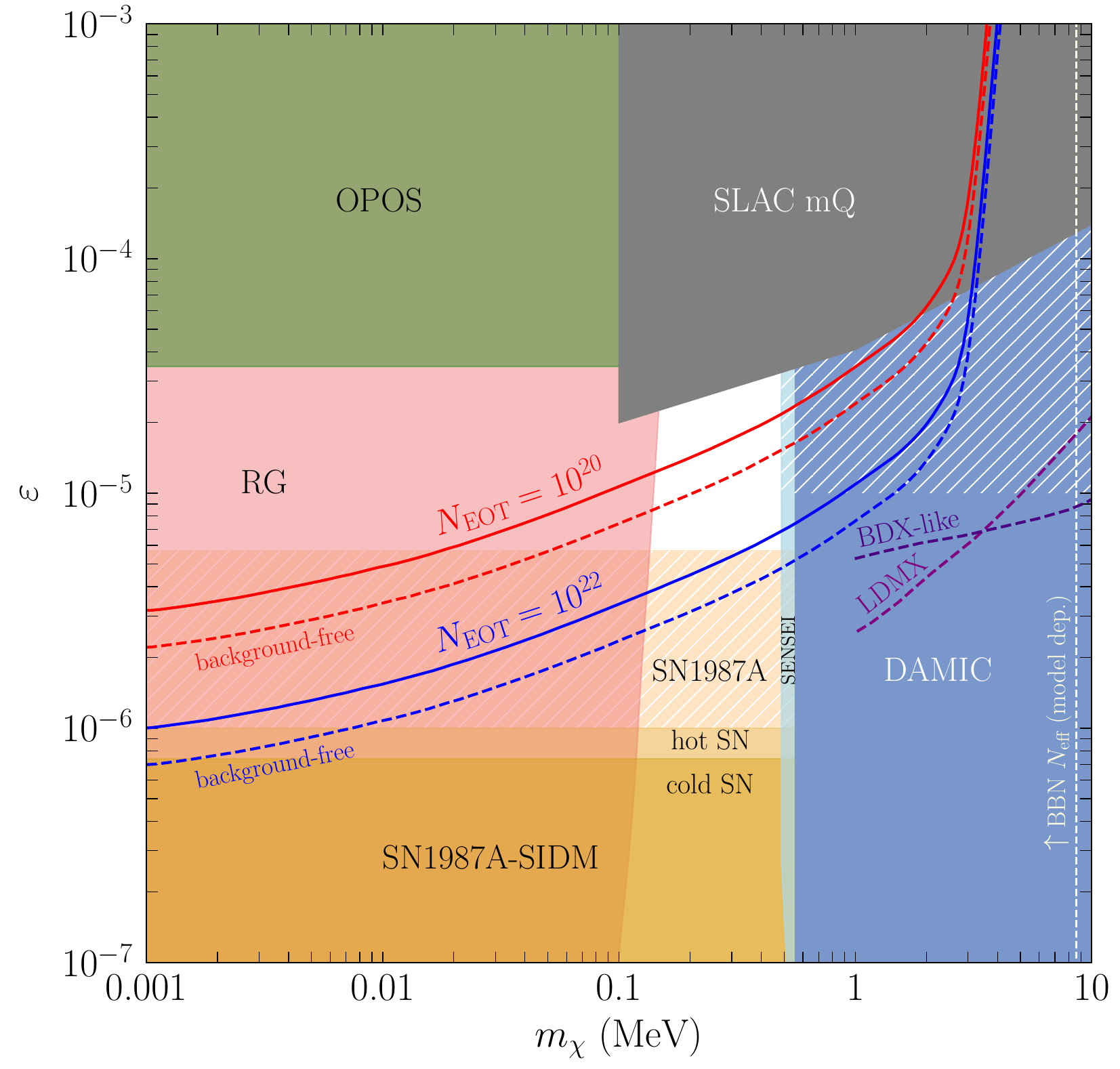}
	\caption{Projected constraints using the set-up described here for a millicharged particle in the 1\,keV to 1\,MeV range for a 100\,MeV beam. The red line shows the exclusion limits for DAMIC-M at $10^{20}$ electrons on target, the blue line for $10^{22}$ electrons on target. The dashed lines describe the background-free case. The open parameter window of millicharge is taken from~\cite{An:2021qdl}. The indirect constraints are from red giants (RG)~\cite{Vogel:2013raa}, SN1987A~\cite{Chang:2018rso} and ortho-positronium decay (OPOS)~\cite{Badertscher:2006fm}. The constraints from SN1987A~\cite{Chang:2018rso} can be weakened up to an order of magnitude when self-interactions of the millicharged particles are considered (SN1987A-SIDM)~\cite{Fiorillo:2024upk} within a hot or cold supernova model, as indicated by the hatching. The direct detection constraints are from SLAC~\cite{Prinz:1998ua}, as well as SENSEI~\cite{SENSEI:2020dpa} and DAMIC~\cite{DAMIC:2019dcn}, which disappear above $\varepsilon\sim10^{-5}$ as indicated by the hatching~\cite{Emken:2019tni}, and are taken from~\cite{An:2021qdl}. The projected bound for LDMX is taken from~\cite{Essig:2024dpa,Berlin:2018bsc} and the projections for a BDX-like set-up combined with CCD detectors is taken from~\cite{Essig:2024dpa}. A cosmological constraint from $N_{\rm eff}$, shown by the dashed vertical line, disfavours the region to the left~\cite{Boehm:2013jpa,Creque-Sarbinowski:2019mcm}; see~\cite{Chu:2023jyb} for scenarios in which this bound can be avoided.}
	\label{fig:milli.window}
\end{figure}

The detector that will be used as a reference is DAMIC-M~\cite{Castello-Mor:2020jhd}. Its active part consists of 50 silicon CCDs. The CCDs are stacked horizontally inside the detector, and it is assumed that the beam line is parallel to the long axis of the CCD stack. The density of detection sites, i.e. the electron density in the CCDs, $n_{\text{det}} $ is the density of silicon. The length of the active stack is $L_{\text{det}} = 50\times 670\,\mu\text{m}$, since the $\chi$ flux is not attenuated due to the smallness of the cross sections involved. The values for the atomic quantities are taken from the Particle Data Group~\cite{ParticleDataGroup:2024cfk}. For the angular aperture of the detector we take $2^{\circ}$, which roughly corresponds to the region in which most $\chi$ particles will be produced, as can be seen in Fig.~\ref{fig:milli.geo} for millicharge and in Fig.~\ref{fig:fm_geo} for EM form factor interactions. This provides a conservative estimate for the number of incoming $\chi$ particles. 

Fig.~\ref{fig:milli.window} shows the DAMIC-M 90\% exclusion limits for the calculation of~(\ref{eq:pdet.6}), using the aforementioned parameters, for a millicharged particle of masses of 1\,keV up to the maximum mass that can be produced by a 100\,MeV beam. The projected limits are obtained by assuming the signal is Poisson distributed, excluding $N_{\rm obs}$ for which $N_{\rm sig} + N_{\rm bkg} \geq N_{\rm obs}$ at a confidence of 90\%, constraining the corresponding coupling. The background data $N_{\rm bkg}$ and the total observed events $N_{\rm obs}$ of DAMIC-M from~\cite{DAMIC-M:2025luv} was used. While Skipper-CCD can detect single charges, the way multiple charge event shapes are registered needs to be modelled for a specific detector. To this end, for simplicity and better generality, we pool the predicted charge bins and the DAMIC-M data. Importantly, as we already only consider events liberating between 2~to~7 charges to account for the sensitivity of Skipper-CCD detectors, as discussed in section 3, this approach should yield a good estimate of the exclusion limits. Since DAMIC-M is located deep underground, the dominant backgrounds are of radioactive origin~\cite{Castello-Mor:2020jhd}. As discussed above, passive shielding between the beam dump and the detector is required to attenuate low-energy SM particles produced in the dump, in particular photons and electrons. High-energy backgrounds originating from the dump are not expected to significantly affect the measurement, as such events deposit substantially larger energies and can be efficiently distinguished from the low-energy signatures relevant here. This discrimination is further enhanced by the Skipper-CCD technology, which provides precise measurement of the liberated charge on an event-by-event basis. Neutrons are also produced in the beam dump; however, existing simulations for beam-dump experiments at even higher energies indicate that their contribution is subdominant. This has been demonstrated both by the BDX collaboration~\cite{BDX:2016akw} and in dedicated studies for a similar set-up with a 10.6\,GeV beam~\cite{Essig:2024dpa}.


We indicate the cosmological sensitivity from big bang nucleosynthesis~\cite{Vogel:2013raa}, CMB effective degrees of freedom~\cite{Boehm:2013jpa}, and 21\,cm~\cite{Creque-Sarbinowski:2019mcm} in Fig.~\ref{fig:milli.window} by a vertical line. The ensuing constraints are model dependent and can be evaded by additional interactions or particle content in the dark sector~\cite{Boehm:2013jpa,Creque-Sarbinowski:2019mcm, Chu:2023jyb}. The constraints derived in this work rely on the direct production and detection of particles effectively interacting with the photon and are therefore insensitive to assumptions about dark-sector dynamics or the cosmological history. Thus, they provide a complementary and robust laboratory probe, independent of cosmological constraints such as relic abundance or effective degrees of freedom.

\begin{figure}%
	\centering
	\begin{subfigure}[h]{0.495\textwidth}
		\centering
		\includegraphics[width=0.99\textwidth, height=!]{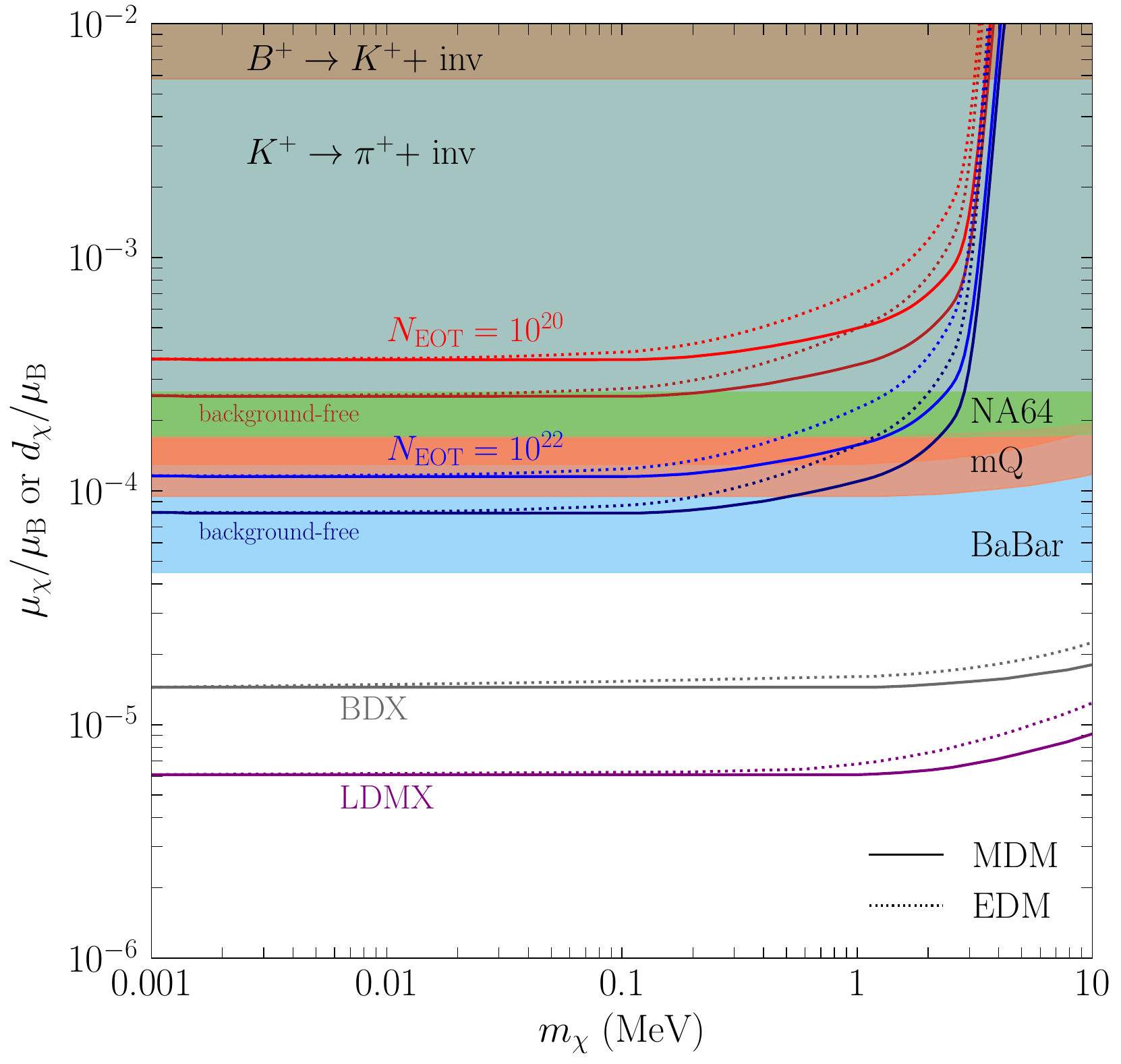}
	\end{subfigure}
	\begin{subfigure}[h]{0.495\textwidth}
		\centering
		\includegraphics[width=0.99\textwidth, height=!]{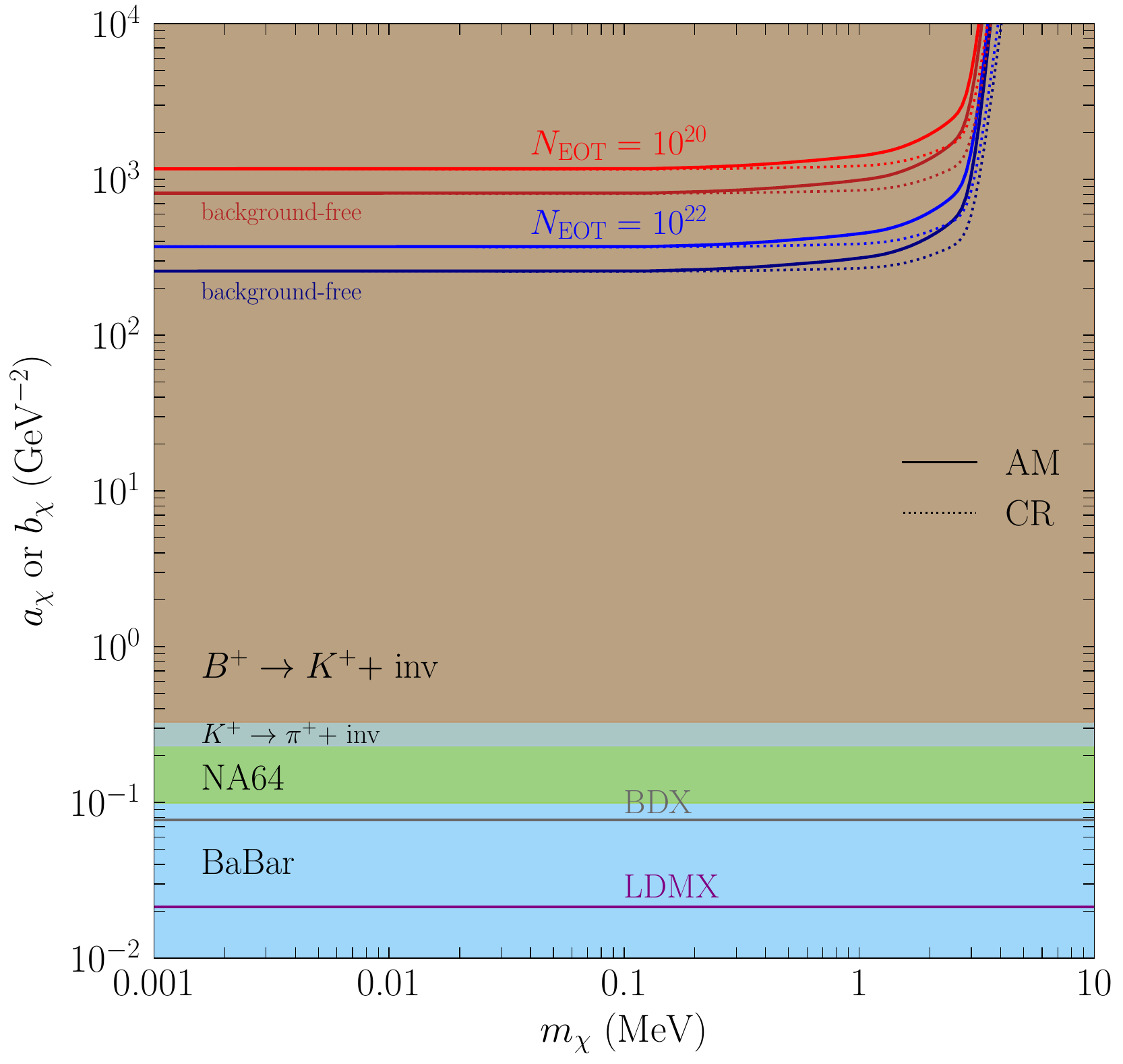}
	\end{subfigure}
	\caption{Projected constraints for MDM and EDM (left) as well as AM and CR (right) for a beam energy of 100\,MeV. The red line shows the 90\% exclusion limits for $10^{20}$ electrons on target, the blue one for $10^{22}$ electrons on target. The constraints set by NA64, BaBar and the predicted bounds for LDMX Phase I and BDX are taken from~\cite{Chu:2018qrm}. The projected limits for Belle-II are not included due to the issues facing the experiment in regards to reaching the integrated luminosity target used in~\cite{Chu:2018qrm} of 50\,ab$^{-1}$~\cite{Natochii:2023thp}.}
	\label{fig:events.dipole}
\end{figure}

The constraints computed for the electromagnetic form factors can be seen in Fig.~\ref{fig:events.dipole}. The exclusion limits are evaluated for a beam energy of 100\,MeV for MDM or EDM in units of the Bohr magneton $\mu_{B}$ and AM and CR in units of GeV$^{-2}$. While the limits of MDM and EDM are comparable, their $\chi$ mass dependence deviates. This difference stems from the $\gamma^5$ that distinguishes MDM and EDM. The left panel of Fig.~\ref{fig:events.dipole} shows that the mass dependence is more pronounced for EDM relative to MDM. The right panel shows the results for AM and CR. Again, as was the case for MDM and EDM, the values of AM and CR are overall similar. However, as before, their mass dependence differs due to the presence of a $\gamma^5$, being more noticeable for AM. 

The choice of a reference beam energy of $100\,\mathrm{MeV}$ is motivated by several considerations. First, as stated in the introduction, the rationale for a theoretical feasibility study at $\mathcal{O}(100\,\mathrm{MeV})$ lies in the comparatively low-energy nature of the experimental setup. While the specific value of $100\,\mathrm{MeV}$ is somewhat arbitrary, Fig.~\ref{fig:prodcross} shows that for a DM mass of $m_{\chi}=1\,\mathrm{MeV}$ a moderate increase in the beam energy to $(200$–$500)\,\mathrm{MeV}$ results in only a 2-5 fold increase in the total production cross section. Such a modest enhancement translates into only a marginal improvement of the $90\%$ exclusion limits derived from Poisson statistics. As illustrated in Fig.~\ref{fig:milli.window}, even a 100-fold increase in the number of electrons on target, which boosts the signal rate by the same factor, leads to a less-than-one-order-of-magnitude improvement in the constraints. Increasing the beam energy to $1\,\mathrm{GeV}$ similarly raises the total production cross section by only about a factor of 6. The increased technical difficulty of achieving higher beam energies, together with the associated rise in backgrounds, therefore offsets the limited gain in sensitivity.
While higher beam energies lead to stronger relativistic beaming and thus improved geometric acceptance, this effect remains modest for energies only moderately above $100\,\mathrm{MeV}$. Moreover, as shown in~\cite{Essig:2024dpa}, even for a $10.6\,\mathrm{GeV}$ beam, the production of millicharged states via meson decays is subdominant in the mass range considered here. Pushing the beam energy significantly beyond $100\,\mathrm{MeV}$ would also require a detailed treatment of nuclear recoil effects, which has already been investigated~\cite{Chu:2018qrm,Essig:2024dpa} and lies beyond the scope of the present work.

Another important result from this work is the analytical expressions derived for a four body phase space in App.~\ref{app:b}. The straightforward integration limits for the $\chi$ particle energy do not hold in the electron-electron bremsstrahlung case. The physical region of $E_{\chi}$ in Fig.~\ref{fig:intlim.physreg} shows a difference in behaviour when considering scattering of two particle with a large separation in mass scale and scattering of identical particles. There is another feature in the physical region arising from the lack of scale separation of the upper mass values considered for $m_{\chi}$ and the scattering partners. Neglecting this change in integration limits leads to significant deviation from the correct results for values of $m_{\chi}$ below the electron mass.

\section{Conclusions}

A novel millicharged particle $\chi$ could conceivably be observed for the described accelerator and CCD-detector set-up. This is especially interesting in the unconstrained window between about 0.1\,MeV and 0.5\,MeV seen in Fig.~\ref{fig:milli.window}. The region not constrained by galactic direct detection experiments seen in Fig.~\ref{fig:milli.window} as the white hatched area, could be probed. Additionally, this can be improved by increasing the electrons on target $N_{\text{EOT}}$. An increase in the beam energy would also lead to an improvement in the constraining power, however this effect is small compared with higher $N_{\text{EOT}}$. When extrapolated to sub-MeV masses, the BDX-like projected limits shown in Fig.~\ref{fig:milli.window} are comparable to ours and could be exceeded by a moderate increase in $N_{\text{EOT}}$. The projected LDMX constraints are somewhat stronger than those of this work if extrapolated.

For EM form factor interactions, however, at a beam energy of 100\,MeV, the predicted events at the detector are well below one for couplings that are not already excluded, as can be seen in Fig.~\ref{fig:events.dipole}. For MDM and EDM it is feasible to increase the beam energy and electrons on target enough to move the projected exclusion limits into a yet un-constrained region. For AM and CR the predictions are too weak to realistically scale this set-up such that it is able to set new limits on the couplings.

While this work was being finalised, we acknowledge a similar set-up was investigated by~\cite{Essig:2024dpa} for millicharged particles. Our work is complementary with their work insofar as they focused on a higher energy beam that allows for production channels other than bremsstrahlung, like meson decay. Furthermore, their work did not treat particles with EM form factor couplings.

\section*{Acknowledgements}
We acknowledge the support of the  Chicago-Vienna Joint Faculty Grant Program ``When Theory Meets Experiment: Dark Matter Detection in CCDs'' and thank Paolo Privitera for suggesting a sensitivity study of underground electron beams. We thank Xiaoyong Chu and Mukul Sholapurkar for their help and acknowledge useful discussions with Stefan Nellen-Mondragón and Garance Lankester-Broche. Funded by the European Union (ERC, NLO-DM, 101044443). This work was also supported by the Research Network Quantum Aspects of Spacetime (TURIS). We acknowledge the financial support by the Vienna Doctoral School in Physics (VDSP). The package \texttt{FeynCalc}~\cite{Mertig:1990an,Shtabovenko:2016sxi,Shtabovenko:2020gxv} was used for algebraic and \texttt{CUBA}~\cite{Hahn:2004fe} for numeric calculations. For cross-checking the results, \texttt{FeynArts}/\texttt{FormCalc}~\cite{Hahn:2000kx,Hahn:1998yk} and \texttt{CalcHep}~\cite{Belyaev:2012qa} were used.

\newpage

\appendix

\section{Analytic Treatment of the Four-Body Phase Space}\label{app:b}

In order to calculate the production cross section at a beam dump, it is necessary to evaluate a four-body phase space integral for the bremsstrahlung-like $\chi$ pair production process ${N(p_1) + e(p_2) \rightarrow X(p_3) + e(p_4) + \chi(p_{\chi}) + \bar{\chi}(p_{\bar{\chi}}) }$, where $N$ is a target with mass $m_N$, X is a final state with inclusive mass $m_X$, $e$ are electrons with $m_e$, with $\chi$ and $\bar{\chi}$ being the new particle pair with mass $m_\chi$. While the electron-electron scattering case will be used in the final calculation, it is still useful to derive the kinematics in full generality with regards to the target particle $N(p_1)$. This allows for greater flexibility when further studying this model beyond the scope of this work. To obtain the following expressions for the electron-electron case, one just substitutes $m_N=m_X=m_e$. When calculating the production cross section
\begin{equation}
d\sigma_{\text{prod}} = \frac{1}{4\sqrt{(p_1\cdot p_2)^2 - m_N^2 m_e^2}} \abs{\mathcal{M}(p_1,p_2\rightarrow p_3,p_4,p_{\chi},p_{\bar{\chi}})}^2 d\Pi_4(p_3,p_4,p_{\chi},p_{\bar{\chi}}) ,
\label{eq:fbps.1} %
\end{equation}
it is necessary to integrate the four-body phase space of the final states
\begin{equation}
d\Pi_4(p_3,p_4,p_{\chi},p_{\bar{\chi}}) = \frac{1}{\mathcal{S}}\prod_{i=3}^{6} \frac{d^3p_i}{(2 \pi)^3 2E_{i}} (2\pi)^4 \delta^{(4)}\Bigg(p_1+p_2-\sum_{j=3}^{6} p_j \Bigg) ,
\label{eq:fbps.2} %
\end{equation}
where $5,6=\chi,\bar{\chi}$ and $\mathcal{S}$ is the symmetry factor that accounts for the over-counting in the final state if $p_3$ and $p_4$ are indistinguishable. To make the integration of this phase space tractable, a common method is phase space splitting, where the system is split into smaller subsystems. First, to introduce these subsystems, one inserts full phase space integrals that evaluate to unity. For the purpose of this work, we seek a differential cross section that depends on the angle between the electron beam line and the particle $\chi$ (and not $\bar{\chi}$ as matter of convention), as well as on the energy of $\chi$. The differential cross section expressed this way is suitable for further calculating the incoming particle flux onto the detector, as it has a limited angular aperture. After introducing these splittings, the original integral is transformed by choosing convenient reference frames for the subsystems.

With this in mind, we introduce the new four-momenta of the subsystems $q_{4\bar{\chi}}\equiv p_4+p_{\bar{\chi}}$, as well as ${q_{34\bar{\chi}}\equiv p_3+p_4+p_{\bar{\chi}}}$. While four outgoing momenta seemingly have 12 degrees of freedom (d.o.f.), this is reduced by four d.o.f. through energy-momentum conservation and by one d.o.f. through the redundancy of rotations around the incoming $\vec{p}_1+\vec{p}_2$ axis~\cite{Byckling}. In total, there are seven independent Lorentz invariants in a system with six four-momenta. It turns out that, for our purposes, a convenient choice of are the five invariants
\begin{gather}
s_{4\bar{\chi}} \equiv (p_4 + p_{\bar{\chi}})^2 , ~ s_{34\bar{\chi}} \equiv (p_3 + p_4 + p_{\bar{\chi}})^2, \nonumber \\
t_{14} \equiv (p_1 - p_4)^2 , ~t_{2\chi} \equiv (p_2 - p_{\chi})^2 , ~ t_{13} \equiv (p_1 - p_3)^2 ,
\label{eq:fbps.3} %
\end{gather}
as well as  the two scalar products $p_2\cdot p_3$ and $p_3\cdot p_4$. This is the same choice as \cite{Chu:2018qrm}. Additionally, the four-vector $q_{23\chi} = p_2 - p_3 - p_{\chi}$ will be used, which has the property 
\begin{equation}
q_{23\chi}^2 \equiv (p_2-p_3-p_{\chi})^2 = m_N^2 + m_{X}^2 + s_{4\bar{\chi}} + t_{2\chi} - s_{34\bar{\chi}} - t_{13} .
\label{eq:fbps.4} %
\end{equation}

For the first sub-frame $\{q_{4\bar{\chi}}, q_{23\chi}, p_3, p_4\}$, we take the rest frame of $\vec{q}_{4\bar{\chi}}$. This will yield an integration over $t_{14}$ and $p_3\cdot p_4$. The integral over the scalar product is expressed through an angle $\cos\varphi_{4}$ in the sub-frame by the relation \cite{Byckling}
\begin{align}
p_3\cdot p_4 &= \frac{ (p_1\cdot p_3) G(p_1, q_{23\chi}; q_{23\chi} ,p_4) }{ -\Delta_2(p_1, q_{23\chi}) } - \frac{ (q_{23\chi}\cdot p_3) G(p_1, q_{23\chi}; p_1 ,p_4) }{ -\Delta_2(p_1, q_{23\chi}) } \nonumber \\
&\phantom{=}\ - \frac{ \sqrt{ \Delta_3(p_1, q_{23\chi}, p_3) \Delta_3(p_1, q_{23\chi}, p_4) } }{ -\Delta_2(p_1, q_{23\chi}) } \cos\varphi_{4} ,
\label{eq:4chisub} %
\end{align}
where ${G(p_1,\dots,p_n;q_1,\dots,q_n)=\det(p_i\cdot q_j)}$ is referred to in~\cite{Byckling} as asymmetric Gram determinant and $\Delta_n(p_1,\dots,p_n)=\det(p_i\cdot p_j) $ is the ordinary Gram determinant. The second sub-frame $\{q_{34\bar{\chi}}, q_{2\chi}, p_2, p_3\}$ in the rest frame of $\vec{q}_{34\bar{\chi}}$ leads to an integral over $t_{13}$ and $p_2\cdot p_3$, again quantified by an angle $\varphi_3$ through
\begin{align}
p_2\cdot p_3 &= \frac{ (p_1\cdot p_2) G(p_1, p_2 - p_\chi; p_2 - p_\chi ,p_3) }{ -\Delta_2(p_1, p_2 - p_\chi) } - \frac{ (m_e^2 - p_2\cdot p_\chi) G(p_1, p_2 - p_\chi; p_1 ,p_3) }{ -\Delta_2(p_1, p_2 - p_\chi) }\nonumber \\
&\phantom{=}\ - \frac{ \sqrt{ \Delta_3(p_1, p_2 - p_\chi, p_2) \Delta_3(p_1, p_2 - p_\chi, p_3) } }{ -\Delta_2(p_1, p_2 - p_\chi) } \cos\varphi_{3} .
\label{eq:34chisub} %
\end{align}
The phase space~(\ref{eq:fbps.2}) is thus transformed into the form used in~(\ref{eq:pcs.1})
\begin{equation}
\frac{d\Pi_{ 4}}{ds_{34\bar{\chi}} dt_{2\chi}} = \frac{1}{\mathcal{S}}ds_{4\bar{\chi}} dt_{13} dt_{14} \frac{d\varphi_{3}}{2\pi} \frac{d\varphi_{4}}{2\pi} \frac{\abs{J}}{(4\pi)^5} \lambda^{-1/2} ( s_{4\bar{\chi}} , m_N^2 , q_{23\chi}^2 ) \lambda^{-1/2} ( s_{34\bar{\chi}} , m_N^2 , t_{2\chi} ) ,
\label{eq:fphi.7} %
\end{equation}
with the Jacobian
\begin{equation}
\frac{d E_{\chi} d\cos\theta_\chi}{ds_{34\bar{\chi}} dt_{2\chi}} = \frac{\abs{J}}{\abs{\vec{p}_{\chi}}} = \frac{1}{4 \abs{\vec{p}_2} \abs{\vec{p}_{\chi}} m_N } .
\label{eq:fphi.6} %
\end{equation}

\begin{figure}%
	\centering
	\includegraphics[width=0.70\textwidth, height=!]{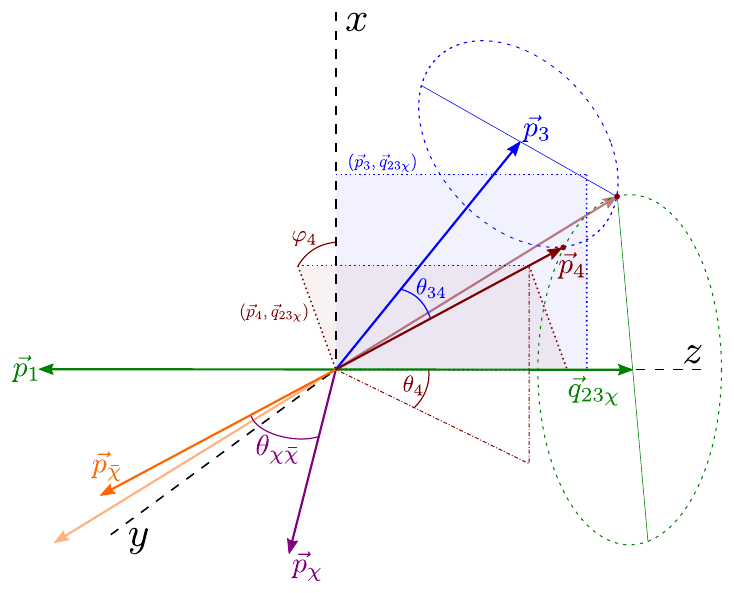}
	\caption{The system of the three-momentum vectors of the phase space in the frame with $\vec{q}_{4\bar{\chi}}=0$. One can clearly see the degeneracy of the angle $\theta_{\chi\bar{\chi}}$, and consequently of $\vec{p}_{\chi}\cdot\vec{p}_{\bar{\chi}}$, illustrated by the lightly coloured vectors next to $\vec{p}_4$ and $\vec{p}_{\bar{\chi}}$. This is due to $\vec{p}_4$ not being uniquely defined by $\vec{q}_{23\chi}$ and $\vec{p}_3$.}
	\label{fig:intlim.momsys}
\end{figure}

In order to calculate the relevant cross sections for this work, we need to express all matrix elements in terms of input parameters $\{m_N,m_X,m_e,m_{\chi},p_1\cdot p_2\}$ and Lorentz invariants $\{s_{34\bar{\chi}}, s_{4\bar{\chi}}, t_{2\chi}, t_{13}, t_{14}, \varphi_{3}, \varphi_{4}\}$ by recasting all possible scalar products of the six four-momenta of the system. There are 15 independent scalar products of six vectors. $p_1\cdot p_2$ is already determined by being an input parameter. Furthermore, seven of the scalar products are directly given through the definition of the Lorentz invariants. The energy-momentum conservation additionally relates six of the scalar products to the others. We notice, that there still remains one seemingly independent scalar product, which can be re-expressed as follows. In 4D, a set of more than four vectors cannot be linearly independent. Hence, it is possible to choose one of the scalar products and determine it through $\det \mathbf{M}=0$, the Gram determinant. We will select $p_{\chi}\cdot p_{\bar{\chi}}$ for this purpose, the same as in~\cite{Chu:2018qrm}. $\mathbf{M}$ is the Gram matrix, which in this case, is a $5\times5$ matrix of scalar products with $(\mathbf{M})_{ij}=p_i\cdot p_j,~~i,j\in{1,3,4,\chi,\bar{\chi}}$. The usage of six vectors here would be redundant, as five are enough for linear dependence. The choice of the five momenta is in principle arbitrary, but here is chosen such that $p_{\chi}\cdot p_{\bar{\chi}}$ appears in the determinant. Unfortunately, the solution to $\det \mathbf{M}=0$ gives two results for $p_{\chi}\cdot p_{\bar{\chi}}$. In order to resolve this degeneracy, we fix the rotation direction of $\varphi_{4}$, see Fig.~\ref{fig:intlim.momsys}, and use both solutions, one for $[0,\pi)$ and one for $(\pi,2\pi]$. 

It is possible to examine this angle degeneracy visually, using Fig.~\ref{fig:intlim.momsys}, in the frame where $\vec{q}_{4\bar{\chi}}=0$. This is equivalent to $\vec{p}_{1}+\vec{q}_{23\chi}=0$, and makes it possible to take $\vec{p}_{3}$ to lie in the $xz$-plane and $\vec{q}_{23\chi}$ to be parallel to the $z$-axis. In this system, the azimuthal angles of $\vec{q}_{4\bar{\chi}}$, and of $\vec{p}_{4}$ and $\vec{p}_{\bar{\chi}}$ individually, have no unique value. Hence, it is necessary to fix the angle of $\vec{p}_{4}$ around the $z$-axis, and therefore the angle around $\vec{q}_{23\chi}$, in order to define $\vec{p}_{\chi}\cdot\vec{p}_{\bar{\chi}}$. However, since the angle between $\vec{p}_{3}$ and $\vec{p}_{4}$, $\theta_{34}$, is already determined in this frame by the invariant $p_3\cdot p_4$, $\vec{p}_{4}$ is confined to a cone around $\vec{p}_{3}$. Additionally, the angle to the $z$-axis of $\vec{p}_{4}$, $\theta_{4}$, is set by the invariant $t_{14}$. In order to satisfy both of these constraints, the azimuthal angle $\varphi_4$ of $\vec{p}_{4}$ must be chosen at a point where its angular freedom around $\vec{p}_{3}$ and $z$-axis intersect. Unfortunately, this still does not completely alleviate the degeneracy of $\vec{p}_{\chi}\cdot\vec{p}_{\bar{\chi}}$, as shown in Fig.~\ref{fig:intlim.momsys}, due to the presence of two intersections. Thus, we resolve this issue, as described in the previous paragraph, by incorporating both solutions, as can be seen in Fig.~\ref{fig:intlim.momsys} indicated by the additional light coloured vectors of $\vec{p}_{4}$ and $\vec{p}_{\bar{\chi}}$.

\subsection{Physical Integration Region for Production}

After reducing the four-body phase space by integrating out the redundant degrees of freedom, we are interested in the physical region of the remaining seven Lorentz invariants. Namely, the integration limits of ${s_{34\bar{\chi}}, s_{4\bar{\chi}}, t_{2\chi}, t_{13}, t_{14}, \varphi_{3}, \varphi_{4}}$. In order to determine these limits, we examine the invariants in the frames used previously to reduce the integration. The limits are then calculated by rewriting them consecutively in terms of input parameters and already determined integration bounds. These limits are 

\begin{equation}
\varphi_{3} \in [0,2\pi), ~ \varphi_{4} \in [0,2\pi) .
\label{eq:intlim.angles} %
\end{equation}
\begin{equation}
(m_X+m_e+m_{\chi})^2 \leq s_{34\bar{\chi}} \leq (\sqrt{s}-m_{\chi})^2 .
\label{eq:intlim.s34c} %
\end{equation}
\begin{equation}
(m_e+m_{\chi})^2 \leq s_{4\bar{\chi}} \leq (\sqrt{s_{34\bar{\chi}}}-m_X)^2 ,
\label{eq:intlim.s4c} %
\end{equation}
\begin{align}
[t_{2\chi}]^{\pm} &= m_e^2 + m_{\chi}^2 - \frac{1}{2 s}(s-m_N^2 + m_e^2)(s + m_{\chi}^2 - s_{34\bar{\chi}}) \nonumber \\
&\pm \frac{1}{2 s} \lambda^{1/2}( s, m_N^2 , m_e^2 )\lambda^{1/2}(s, s_{34\bar{\chi}} , m_{\chi}^2 ) ,
\label{eq:intlim.t2c} %
\end{align}
\begin{align}
[t_{13}]^{\pm} &= m_N^2 + m_X^2 - \frac{1}{2 s_{34\bar{\chi}}}(s_{34\bar{\chi}}+m_N^2 - t_{2\chi} )(s_{34\bar{\chi}} + m_X^2 - s_{4\bar{\chi}}) \nonumber \\ 
&\pm \frac{1}{2 s_{34\bar{\chi}}} \lambda^{1/2}( s_{34\bar{\chi}}, t_{2\chi}, m_N^2 )\lambda^{1/2}(s_{34\bar{\chi}}, s_{4\bar{\chi}} , m_X^2 ) ,
\label{eq:intlim.t13} %
\end{align}
\begin{align}
[t_{14}]^{\pm} &= m_N^2 + m_e^2 - \frac{1}{2 s_{4\bar{\chi}}}(s_{4\bar{\chi}}+m_N^2 - q_{23\chi}^2 )(s_{4\bar{\chi}} + m_e^2 - m_{\chi}^2 ) \nonumber \\ 
&\pm \frac{1}{2 s_{4\bar{\chi}}} \lambda^{1/2}( s_{4\bar{\chi}}, q_{23\chi}^2, m_N^2 )\lambda^{1/2}(s_{4\bar{\chi}}, m_{\chi}^2 , m_e^2 ) .
\label{eq:intlim.t14} %
\end{align}

The previously derived limits are frame-independent, as they solely consist of Lorentz invariant quantities. In this work, however, we are explicitly interested in the lab frame of the target-beam system. Therefore, the differential variables $s_{34\bar{\chi}}$ and $t_{2\chi}$ in~(\ref{eq:fphi.7}) need to be replaced by the energy of the outgoing $\chi$ particle $E_{\chi}$ and the cosine of the $\chi$ particle's angle to the beam line, $\cos\theta_{\chi}$. In the lab frame $\vec{p}_1=0$, the incoming energy is $s=m_N^2+m_e^2+2m_N E_2$ and the $\chi$ energy is 
\begin{equation}
E_{\chi} = \frac{1}{2}m_N + E_2 + \frac{1}{2 m_N} (t_{2\chi} -s_{34\bar{\chi}}).
\label{eq:intlim.Echi} %
\end{equation}
One has to be careful when translating the limits of the Lorentz invariants into the lab frame kinematic variables. On first look the physical region of $E_\chi$ is determined by combining the limits of $s_{34\bar{\chi}}$ and $t_{2\chi}$ such that they minimise or maximise $E_\chi$. This yields for the integration region
\begin{align}
E_{\chi} &\geq \frac{m_{\chi} ( m_N + E_2 ) }{ \sqrt{ m_N^2 + m_e^2 + 2m_N E_2 }} \label{eq:intlim.Echi.lim_low} \\
E_{\chi} &\leq \frac{1}{2}\left(m_N+E_2\right) + \frac{(m_N + E_2)[ m_{\chi}^2 - (m_X+m_e+m_{\chi})^2]}{2(m_N^2 + m_e^2 + 2m_N E_2 )} \nonumber \\
& + \frac{\sqrt{E_2^2-m_e^2}}{2(m_N^2 + m_e^2 + 2m_N E_2 )}\sqrt{m_N(m_N+2E_2)-m_X(m_X+2m_e)}\nonumber\\
&\phantom{=}\ \times \sqrt{m_N(m_N+2E_2)-(m_X+2m_{\chi})(m_X+2(m_{\chi}+m_e))} .
\label{eq:intlim.Echi.lim} %
\end{align}
\begin{figure}%
	\centering
	\includegraphics[width=0.7\textwidth, height=!]{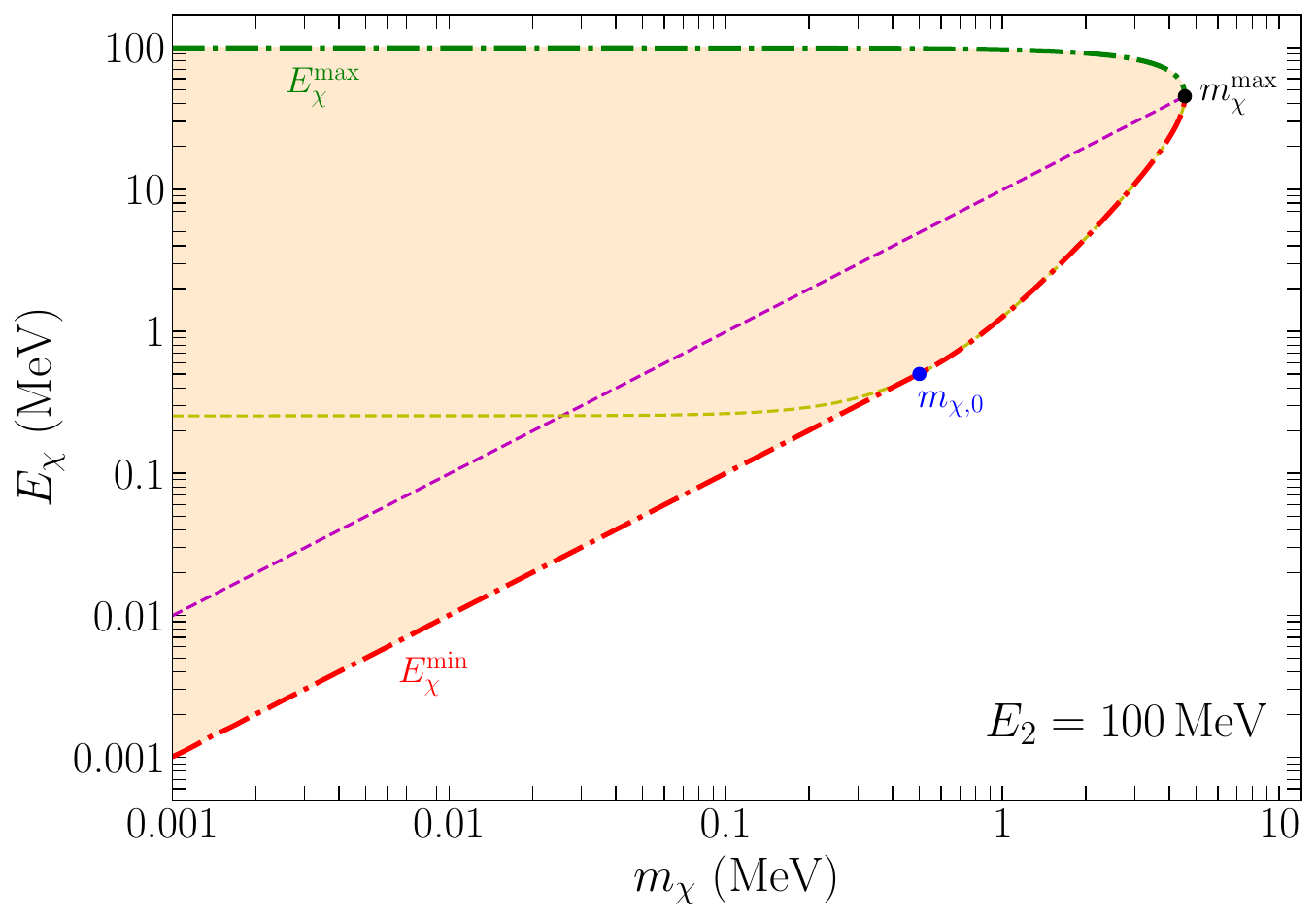}
	\caption{The physical region of the energy $E_\chi$ of the final state $\chi$ particle shown as the orange shaded area for an incoming electron's energy $E_2$ of 100\,MeV. This region is bounded from above by the green line, corresponding to the upper limit~(\ref{eq:intlim.Echi.lim}) with $m_N=m_X=m_e$, and bounded from below by the red line. Importantly, the changing behaviour of the minimum at  $m_{\chi ,0}$ is shown. The magenta line represents the form of the minimum for the nuclear case~(\ref{eq:intlim.Echi.lim_low}). The yellow line corresponds to~(\ref{eq:intlim.echimin}) and shows the matching with the red line until it reaches $m_{\chi ,0}$, after which they significantly diverge.}
	\label{fig:intlim.physreg}
\end{figure}
\eqref{eq:intlim.Echi.lim_low} is correct for $m_\chi < m_N$ and $m_e\ll m_N$, when the incoming electron's mass can be neglected and the mass scales are clearly separated. Importantly, however, in the case of electron-electron bremsstrahlung together with $m_\chi$ in the keV to MeV range, the electron mass cannot be neglected.$^{\ref{fn:1}}$ Thus, for the process investigated in this work, the different particles have no clear scale separation, resulting in a change of the behaviour of the lower bound of $E_\chi$. The focus will now be on the $e^- e^-$ scattering case and thus $m_N=m_e$ and $m_X=m_e$. From the kinematic limit of $m_\chi$ production
\begin{equation}
m_{\chi}^{\text{max}} = \sqrt{\frac{m_e(m_e+E_2)}{2}} - me
\label{eq:intlim.mchimax} %
\end{equation}
up to a critical value 
\begin{equation}
m_{\chi ,0} = m_e \frac{E_2-m_e}{E_2+3m_e}
\label{eq:intlim.mchicrit} %
\end{equation}
at which $\chi$ can be produced at rest, the lower bound of the energy is 
\begin{align}
E_{\chi}^{\text{min}} &= \frac{E_2 + m_e}{2} + \frac{1}{4m_e} \left[ m_{\chi}^2 - (2m_e+m_{\chi})^2 \right] \nonumber \\
&- \frac{\sqrt{E_2^2-m_e^2}}{4m_e(m_e+E_2)}\sqrt{2m_e (E_2 -m_e)}\sqrt{2m_e E_2-2(m_e^2+2m_{\chi}^2+4m_e m_{\chi})}.
\label{eq:intlim.echimin} %
\end{align}
For $m_{\chi} < m_{\chi ,0}$, the lower bound is $E_{\chi}^{\text{min}}=m_\chi$. The physical region of $E_\chi$ at different values of $m_{\chi}$ is shown in Fig.~\ref{fig:intlim.physreg} for $E_2=100$\,MeV. Expectedly, for an observable like the energy of $\chi$, the bounds of the physical region are solely determined by the physical input parameters, i.e. the total energy we put into the system and the masses of the participating particles. 

For $\cos\theta_{\chi}$, we need to carefully examine the boost from the centre-of-mass (CM) frame $\vec{p}_1+\vec{p}_2=0$, in which the limits of $t_{2\chi}$ are defined with the CM angle $\cos\theta_{2\chi}=\pm1$, into the lab frame where $\vec{p}_1=0$. In the lab frame it is convenient to put the incoming momentum $\vec{p}_2$, i.e. the beam line, into the z-axis, thus having $t_{2\chi}= m_e^2 + m_{\chi}^2 - 2E_2 E_{\chi} + 2 \abs{\vec{p}_2}\abs{\vec{p}_{\chi}} \cos\theta_{\chi}$. This allows for the reduction of the boost from the CM frame to the lab frame into a transformation along the z-axis by aligning the CM three-momenta $\vec{p}_{1,\text{CM}}=-\vec{p}_{2,\text{CM}}$ parallel to this axis. Hence, the maximum value for the $\chi$'s angle is evidentially $0^\circ$, or $\cos\theta_{\chi}=1$, in the lab frame as well. This is because the boost from CM to lab frame is along the beam line, so a vector parallel to it will experience no change in direction. In principle, the lowest value of $\cos\theta_{\chi}$ could be~$-1$. This, however, will be increased by the velocity of $\chi$. The minimum value of $\cos\theta_{\chi}$ is then determined by the minimum value of $s_{34\bar{\chi}}$ for a fixed $E_{\chi}$. To find the exact lower limit, we use the definition of the invariant $s_{34\bar{\chi}}$ to derive a form for the cosine of the angle as
\begin{equation}
\cos\theta_{\chi} = \frac{s_{34\bar{\chi}} - m_N^2 - m_e^2 - m_{\chi}^2 + 2E_2 E_{\chi} + 2m_N E_{\chi} -2m_N E_{2} }{2\sqrt{E_{\chi}^2-m_{\chi}^2}\sqrt{\vphantom{E_{\chi}^2}E_{2}^2-m_{e}^2}} .
\label{eq:intlim.cosTH} %
\end{equation}
After inserting the minimum value of $s_{34\bar{\chi}}$, $(m_X+m_e+m_{\chi})^2$, into~(\ref{eq:intlim.cosTH}) and keeping in mind the fact that the absolute minimum value of $\cos\theta_{\chi}$ is $-1$, the integration limits are
\begin{equation}
\begin{split}
1\geq \cos\theta_{\chi} \geq \max\Bigg\{ -1,& \frac{E_2 (E_{\chi}-m_N)}{\sqrt{E_{\chi}^2-m_{\chi}^2}\sqrt{ \vphantom{E_{\chi}^2} E_{2}^2-m_{e}^2} } \\
&+\frac{m_N(2E_{\chi}-m_N)+m_X(2m_{\chi}+m_X)+2m_e(m_X+m_{\chi})}{2\sqrt{E_{\chi}^2-m_{\chi}^2}\sqrt{\vphantom{E_{\chi}^2} E_{2}^2-m_{e}^2}} \Bigg\}.
\end{split}
\label{eq:intlim.cosTH.lim} %
\end{equation}
In the case of $e^- e^-$ bremsstrahlung, the bounds are~(\ref{eq:intlim.cosTH.lim}) with $m_N=m_X=m_e$.

\section{Feynmann-Rules, Detection/Production Cross Sections}\label{app:a}

The interactions~(\ref{eq:mod.1a}-\ref{eq:mod.1e}) have the following vertex functions 
\begin{subequations}
	\label{eq:mod.3}
	\begin{alignat}{2}
	\label{eq:mod.3a}
	\text{$\varepsilon e$:}& &i \Gamma_{\varepsilon}^{\mu} (q) &= i \varepsilon e \gamma^{\mu},
	\\
	\label{eq:mod.3b}
	\text{MDM:}& &i \Gamma_{\text{M}}^{\mu} (q) &= - \mu_\chi \sigma^{\mu\nu} q_\nu,
	\\
	\label{eq:mod.3c}
	\text{EDM:}& &i \Gamma_{\text{E}}^{\mu} (q) &= - i d_\chi \sigma^{\mu\nu} \gamma^5 q_\nu ,
	\\
	\label{eq:mod.3d}
	\text{AM:}& &i \Gamma_{\text{A}}^{\mu} (q) &= - i a_\chi (q^2 \gamma^\mu - q^\mu \slashed{q} ) \gamma^5 ,
	\\
    	\label{eq:mod.3e}
	\text{CR:}& &i \Gamma_{\text{C}}^{\mu} (q) &= i b_\chi (q^2 \gamma^\mu - q^\mu \slashed{q} ),
	\end{alignat}
\end{subequations}
where the $q$ refers to the momentum of the photon at the vertex. Using these, the cross section for free elastic 2-to-2 scattering of $\chi$ on electrons can be evaluated for this model. This will be done in the lab frame, i.e. the rest frame of the target electron. 

\subsection{Detection Cross Section}

\begin{figure}[H]
	\centering
	\begin{tikzpicture}
	\node[xshift=-10] (0,0) {$\chi$};
	\draw[-{Latex}, line width=0.3mm,xshift=-5] (0,0) -- node[label=below:$p_1$]{} (1.5,0);
	\draw[dashed, line width=0.3mm]  (1.5,0) node[circle,fill=black,scale=0.5,label=$e^-$,label=below:$p_2$]{} -- (2.9,0) ;
	\draw[-{Latex}, line width=0.3mm,xshift=3,yshift=3] (1.5,0) -- node[label=$p_3$]{} (2.8,1) node[label=right:$\chi$,xshift=-4]{};
	\draw[-{Latex}, line width=0.3mm, xshift=3,yshift=-3] (1.5,0) -- node[label=below:$p_4$]{} (2.8,-1) node[label=right:$e^-$,xshift=-4]{};
	\draw[domain=325:360,line width=0.2mm] plot[shift={(1.4,0)}] ({cos(\x)}, {sin(\x)});
	\node[xshift=-10] at (2.55,-0.27) {$\theta$};
	\end{tikzpicture}
	\caption{Kinematics of $e^-\chi$ elastic scattering in the electron's rest frame with corresponding momenta. The scattering angle of the electron is denoted by $\theta$ in the diagram.}
	\label{fig:echiES.kin}
\end{figure}
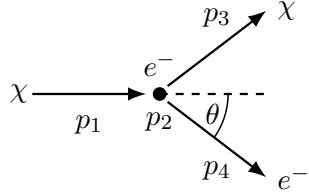

The detection cross section of the $\chi$-particles produced in the potential experimental set-up described is, to first order, this elastic cross section. The conventions and relativistic kinematics can be seen in Fig.~\ref{fig:echiES.kin}. The outgoing particles have four-momenta $p_{3,4}=(E_{3,4},\vec{p}_{3,4})^{\text{T}}$, the incoming $\chi$-particle has a four-momentum $p_1 = (E_{\chi},\vec{p}_1)^{\text{T}}$, while the initial electron has $p_2=(m_e,\vec{0})^{\text{T}}$ in the lab frame. The recoil cross section of the scattering is needed to relate measurements of the energy, deposited by the $\chi$ particle into the electrons of a detector, to the parameters of the model. Hence we define the scattered electron's relativistic recoil energy as
\begin{equation}
E_R = E_4 - m_e .
\label{eq:echiES.1} %
\end{equation}
The recoil cross section in the lab frame is
\begin{equation}
 \frac{d\sigma_{e^-\chi}}{dE_R}  = \frac{\abs{\mathcal{M}}^2}{32\pi m_e (E_{\chi}^2-m_\chi^2)}.
\label{eq:echiES.8} %
\end{equation}
The differential recoil cross section~(\ref{eq:echiES.8}) with the fine-structure constant $\alpha = e^2/(4\pi)$:
\begin{subequations}
	\label{eq:echiES.10}
	\begin{alignat}{2}
	\label{eq:echiES.10a} %
	\left( \frac{d\sigma_{e\chi}}{dE_R} \right)_{\varepsilon e} &= \frac{\pi \varepsilon^2 \alpha^2}{m_e^2 (E_{\chi}^2 - m_\chi^2) E_R^2} \Big[ 2m_e E_{\chi}^2 + m_e E_R( E_R - m_e - 2E_{\chi} ) - m_\chi^2 E_R \Big] ,
	\\
	\label{eq:echiES.10b} %
	\left( \frac{d\sigma_{e\chi}}{dE_R} \right)_{\text{MDM}} &= \frac{\alpha \mu_{\chi}^2}{2 m_e (E_{\chi}^2 - m_\chi^2) E_R} \Big[ 2m_e (E_{\chi}^2 - E_{\chi} E_R - m_\chi^2) + m_\chi^2 E_R \Big] ,
	\\
	\label{eq:echiES.10c} %
	\left( \frac{d\sigma_{e\chi}}{dE_R} \right)_{\text{EDM}} &= \frac{\alpha d_{\chi}^2}{2 m_e (E_{\chi}^2 - m_\chi^2) E_R} \Big[ 2m_e E_{\chi}^2 -2m_e E_{\chi} E_R - m_\chi^2 E_R \Big] ,
	\\
	\label{eq:echiES.10d} %
	\left( \frac{d\sigma_{e\chi}}{dE_R} \right)_{\text{AM}} &= \frac{\alpha a_{\chi}^2}{(E_{\chi}^2 - m_\chi^2)} \Big[ m_e (2E_{\chi}^2 - 2E_{\chi} E_R +E_R^2 - 2m_\chi^2) + (m_\chi^2 - m_e^2)E_R  \Big]  ,
	\\
	\label{eq:echiES.10e} %
	\left( \frac{d\sigma_{e\chi}}{dE_R} \right)_{\text{CR}} &= \frac{\alpha b_{\chi}^2}{(E_{\chi}^2 - m_\chi^2)} \Big[ m_e (2E_{\chi}^2 - 2E_{\chi} E_R +E_R^2) - (m_\chi^2 + m_e^2)E_R  \Big] .
	\end{alignat}
\end{subequations}

In order to evaluate the total recoil cross section, it is necessary to determine the integration limits for $E_R$. Evidently, the theoretically possible minimum recoil energy is $E_R=0$, meaning no scattering occurred and the electron acquires no kinetic energy. Additionally, this also represents the limiting case for the angle in Fig.~\ref{fig:echiES.kin}, namely $\theta = 90^\circ$, at which no momentum transfer occurs. The recoil energy reaches its maximum when backscattering off of the incoming $\chi$-particle, i.e. at $\theta = 0$. The maximal energy transfer is thus
\begin{equation}
E_R^{\text{max}} = \frac{2 m_e (E_{\chi}^2 - m_\chi^2)}{m_e (2E_{\chi} +m_e) + m_\chi^2} .
\label{eq:echiES.11} %
\end{equation}

As described in Section 3, for the calculation of the bound-electron cross section the dielectric function $\epsilon(E_{R},k)$ of silicon is needed. While the imaginary part $\epsilon_2(E_{R},k)$ is calculated via a DFT method in order to achieve better results for high momentum transfer~\cite{Essig:2024ebk}, the absolute value $|\epsilon(E_{R},k)|^2$ needs to be supplied from another source. Following~\cite{Essig:2023wrl}, $|\epsilon(E_{R},k)|^2$ is obtained from the Lindhard model using only measured data as input. The dielectric function~\cite{Essig:2023wrl,Dressel_Gruener_2002} is given by
\begin{equation}
\epsilon_{\rm Lind}(E_{R},k) = 1 + \frac{3\omega_p^2}{k^2v_F^2}\left(\frac{1}{2} + F_+ + F_- \right),
\label{eq:lindhard} %
\end{equation} %
where $\omega_p^2=4\pi e^2 n_e/m_e$ is the plasma frequency of silicon, with $n_e$ being the electron density of the valence band of silicon, $k_F = \sqrt{3\pi^2} n_e$ is the corresponding Fermi-momentum and $v_F=k_F/m_e$ is the Fermi velocity in the Lindhard model. The $F_{\pm}$ are defined as
\begin{align}
F_{\pm} &= \frac{k_F}{4 k} (1 - Q_{\pm}^2)\operatorname{Ln}{\left( \frac{Q_{\pm}+1}{Q_{\pm}-1} \right)}, \nonumber \\
Q_{\pm} &= \frac{k}{2 k_F} \pm \frac{m_e}{k_F k} (E_R + i\Gamma_{p})
\label{eq:lindhard.extra} %
\end{align}
where Ln denotes the principal branch of the complex logarithm and $\Gamma_{p}$ is the width of the plasmon peak, here $\Gamma_{p}\sim0.1\omega_p$~\cite{Essig:2023wrl}.

\subsection{Production Cross Section}\label{app:c}

\begin{figure}%
	\centering
	\begin{subfigure}[h]{0.45\textwidth}
		\centering
		\includegraphics[width=0.67\textwidth, height=!]{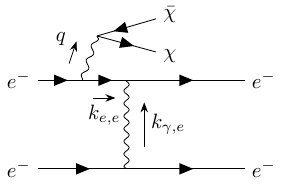}
	\end{subfigure}
	\begin{subfigure}[h]{0.45\textwidth}
		\centering
		\includegraphics[width=0.67\textwidth, height=!]{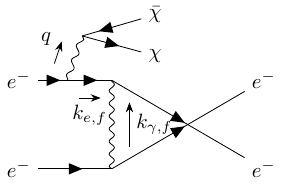}
	\end{subfigure} %
	\begin{subfigure}[h]{0.45\textwidth}
		\centering
		\includegraphics[width=0.67\textwidth, height=!]{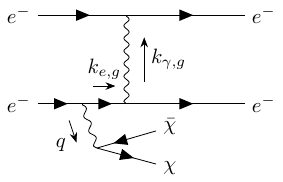}
	\end{subfigure} %
	\begin{subfigure}[h]{0.45\textwidth}
		\centering
		\includegraphics[width=0.67\textwidth, height=!]{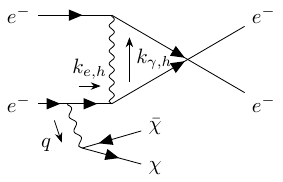}
	\end{subfigure} %
	\caption{Production of $\chi\bar{\chi}$ pairs from a photon radiated by an incoming electron. The $k_{\gamma,n}$ and $k_{e,n}$, with $n\in\{e,f,g,h\}$, represent the momenta for each diagram indexed by $n$.}
	\label{fig:prodFeyn.i} %
\end{figure}

As can be seen from the Feynman diagrams in Fig.~\ref{fig:prodFeyn.f} and Fig.~\ref{fig:prodFeyn.i}, it is possible to split the production matrix element $i\mathcal{M}_{\text{prod}}$ into an electron scattering part with the emission of a virtual photon $i\mathcal{M}_{ee}^{\alpha}$, and a $\chi$ particle-antiparticle pair production part $i\mathcal{M}_{\chi,\alpha}$, as
\begin{equation}
i\mathcal{M}_{\text{prod}} = (i\mathcal{M}_{ee}^{\alpha}) \frac{-i g_{\alpha\beta}}{q^2+i\epsilon} (i\mathcal{M}_{\chi}^{\beta}) = \frac{-i }{q^2+i\epsilon} (i\mathcal{M}_{ee}^{\alpha}) (i\mathcal{M}_{\chi,\alpha} ) .
\label{eq:prodFeyn.1} %
\end{equation}
Again we use Feynman gauge, as this does not affect tree level calculation due to the on-shell spinors. The pair production part is then 
\begin{equation}
i\mathcal{M}_{\chi,\alpha}(q) = \bar{u}_{\chi} (i\Gamma_{\alpha}(q)) v_{\bar{\chi}},
\label{eq:prodFeyn.2} %
\end{equation}
where $u_i$ refers to the spinor for a fermion with momentum and mass $(p_i,m_i)$ and its corresponding spin index, $v_j$ refers to the spinor for an antifermion with momentum and mass $(p_j,m_j)$ and its corresponding spin index, while $i\Gamma_{\alpha}(q)$ refers to the EM form factor vertices in~(\ref{eq:mod.3a}-\ref{eq:mod.3e}). 
The electron-electron scattering part is 
\begin{equation}
i\mathcal{M}_{ee}^{\alpha} = \sum_{n=1}^{8} i\mathcal{M}_{ee, n}^{\alpha} = \sum_{\text{out: } m=1}^{4} i\mathcal{M}_{ee, m}^{\alpha} + \sum_{\text{in: } r=1}^{4} i\mathcal{M}_{ee, r}^{\alpha} , 
\label{eq:prodFeyn.3} %
\end{equation}
where the first part of the sum are the four diagrams in Fig.~\ref{fig:prodFeyn.f} and the second part of the sum are the diagrams in Fig.~\ref{fig:prodFeyn.i}. The general form of the amplitude for the case of a photon radiating from the outgoing electron is
\begin{equation}
i\mathcal{M}_{ee,m}^{\alpha} = (ie)^3 \bar{u}_k \gamma^{\mu} u_i \frac{-i g_{\mu\nu}}{k_{\gamma,m}^2+i\epsilon} \bar{u}_l \gamma^{\alpha} \frac{i (\slashed{k}_{e,m} +m_e)}{k_{e,m}^2-m_e^2+i\epsilon} \gamma^{\nu} u_j ,
\label{eq:prodFeyn.4} %
\end{equation}
with the corresponding electrons $\{i,j,k,l\}$, while $k_{\gamma,m}$ and $k_{e,m}$ refer to the momenta for each diagram as denoted in Fig.~\ref{fig:prodFeyn.f} for $m\in\{a,b,c,d\}$. Similarly, the amplitude for a photon radiating from the incoming electron is
\begin{equation}
i\mathcal{M}_{ee,r}^{\alpha} = (ie)^3 \bar{u}_k \gamma^{\mu} u_i \frac{-i g_{\mu\nu}}{k_{\gamma,r}^2+i\epsilon} \bar{u}_l \gamma^{\nu} \frac{i (\slashed{k}_{e,r} +m_e)}{k_{e,r}^2-m_e^2+i\epsilon} \gamma^{\alpha} u_j ,
\label{eq:prodFeyn.5} %
\end{equation}
where $k_{\gamma,r}$ and $k_{e,r}$ are the momenta seen in Fig.~\ref{fig:prodFeyn.i} for $r\in\{e,f,g,h\}$. We calculate the spin-summed squared matrix elements from these amplitudes using FeynCalc in Mathematica. Hence, the final tree-level form for the electron-electron $\chi$ pair production cross section~(\ref{eq:pcs.1}) is obtained by integrating the analytic expressions of the squared amplitude and phase space.

\section{Results for other EM Form Factor Interactions}

The values for the reference cross section $\hat{\sigma}_{\rm tot}$  used in Fig.~\ref{fig:prodcross} are listed in Tab.~\ref{tab:ref}. The left column shows $\hat{\sigma}_{\rm tot}$  for fixed $\chi$-mass and the right column for fixed input energy.
The values of the coupling constants used are not yet excluded for the considered mass range. In Fig.~\ref{fig:prodcross} these drop out. The scaling of the total cross section with coupling constants is elementary,  such that $\sigma_{\rm tot}\sim\alpha_c^4$, where $\alpha_c$ is stands for either millicharge or an EM form factor. Table~\ref{tab:ref} shows that the mass-dependence for small $m_{\chi}$ between operators that only differ by $\gamma^5$ vanishes. 

\begin{table}[tb]
	\captionsetup{font=small}
	\caption{The reference cross sections $\hat{\sigma}_{\rm tot}$ for the different form factors in Fig~\ref{fig:prodcross}. The left column shows the references for the mass dependence of $\sigma_{\rm tot}$, while the right column shows them with respect to the incoming energy dependence of $\sigma_{\rm tot}$. }
	\centering
	\begin{tabular}{c c c}
		\toprule  
		& \multicolumn{2}{c}{ $\hat{\sigma}_{\rm tot}$ (cm$^2$MeV$^{-1}$)} \\
		\cmidrule(lr){2-3}
		Form Factor & $\hat{\sigma}_{\rm tot}(m_{\chi,\rm ref}=1\,\text{keV})$ & $\hat{\sigma}_{\rm tot}(E_{2,\rm ref}=10\,\text{MeV})$ \\
		\midrule
		$\varepsilon e$, $\varepsilon=10^{-5}$ & $4.1 \times10^{-38}$ & $1.6 \times10^{-40}$ \\
		MDM, $\mu_{\chi}=10^{-6}\mu_B$& $7.9 \times10^{-42}$  & $1.6 \times10^{-41}$   \\
		EDM, $d_{\chi}=10^{-6}\mu_B$ & $7.9 \times10^{-42}$  & $9.4 \times10^{-42}$   \\
		AM, $a_{\chi}=10^{-3}\text{GeV}^{-2}$ & $7.4 \times10^{-46}$ & $1.7 \times10^{-44}$  \\
        CR, $b_{\chi}=10^{-3}\text{GeV}^{-2}$ & $7.4 \times10^{-46}$  & $1.9 \times10^{-44}$ \\
		\bottomrule
	\end{tabular}
    \label{tab:ref}
\end{table}

\begin{figure}%
	\centering
	\begin{subfigure}[h]{0.495\textwidth}
		\centering
		\includegraphics[width=0.99\textwidth, height=!]{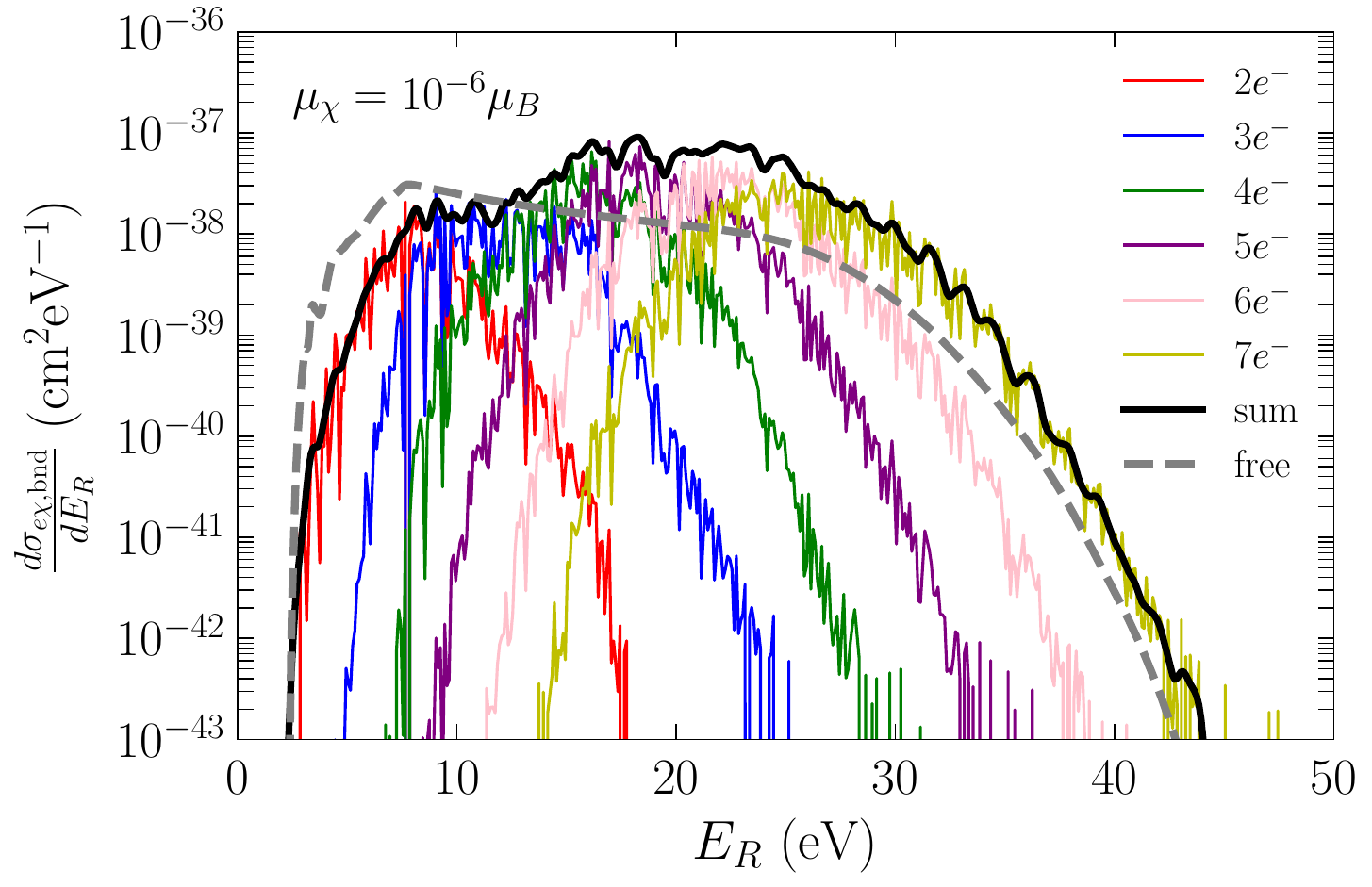}
	\end{subfigure}
	\begin{subfigure}[h]{0.495\textwidth}
		\centering
		\includegraphics[width=0.99\textwidth, height=!]{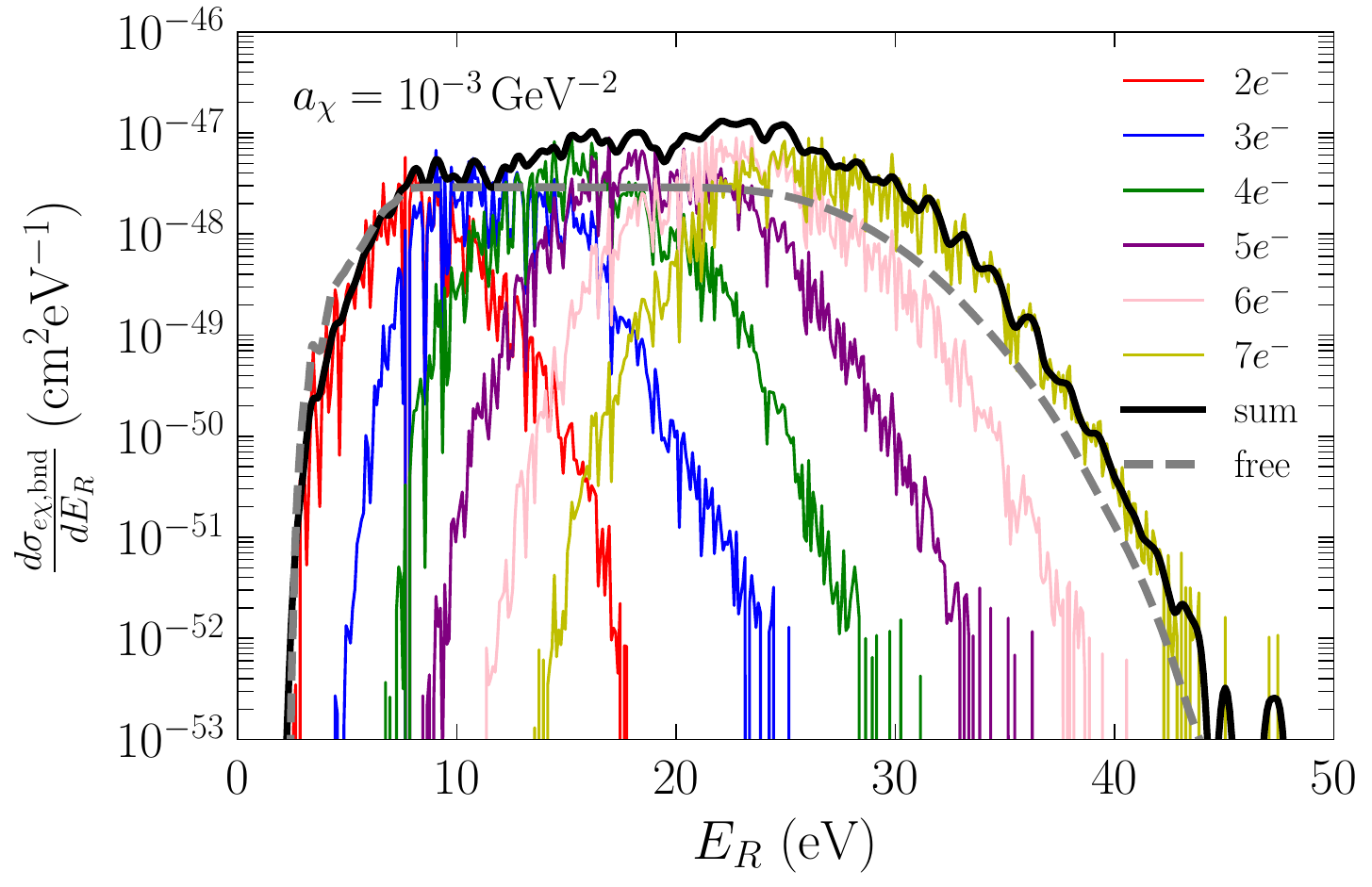}
	\end{subfigure}
	\caption{Differential cross section for the creation of two to seven charges with respect to the recoil energy deposited into electrons by $\chi$ of energy $E_{\chi}=5$\,MeV and mass $m_{\chi}=1$\,keV. The \emph{left panel} shows the values for MDM and the \emph{right panel} shows AM. The results are obtained using an ab-initio electron-loss function for silicon~\cite{Dreyer:2023ovn}. The black line shows the Gaussian-smoothed sum of the bins. For the higher mass-dimension operators shown here (in comparison to millicharge) the peak at the plasmon resonance frequency is less pronounced. This is due to increased contributions at higher energies for the effective couplings. Shown as a dashed grey line is the respective free case.}
	\label{fig:binned_comp_fm}
\end{figure}

In analogy to the right panel of Fig.~\ref{fig:binned_comp}, Fig.~\ref{fig:binned_comp_fm} shows the bound electron scattering cross section for the EM form factors as labeled. The left panel of Fig.~\ref{fig:binned_comp_fm} depicts the dimension-5 operators, while the right panel shows the dimension-6 operator. The difference between MDM and EDM, or AM and CR, vanishes in the semi-classical cross section~\eqref{eq:pdet.ss}, as can be seen from the identical form of $F(E_{R},k)$ in~\eqref{eq:pdet.ffb} and~\eqref{eq:pdet.ffd}. One can notice that compared to millicharge in Fig.~\ref{fig:binned_comp}, the enhancement at the plasmon peak over the free scattering case decreases with increasing mass-dimension of the operators.

\begin{figure}%
	\centering
	\begin{subfigure}[h]{0.495\textwidth}
		\centering
		\includegraphics[width=0.99\textwidth, height=!]{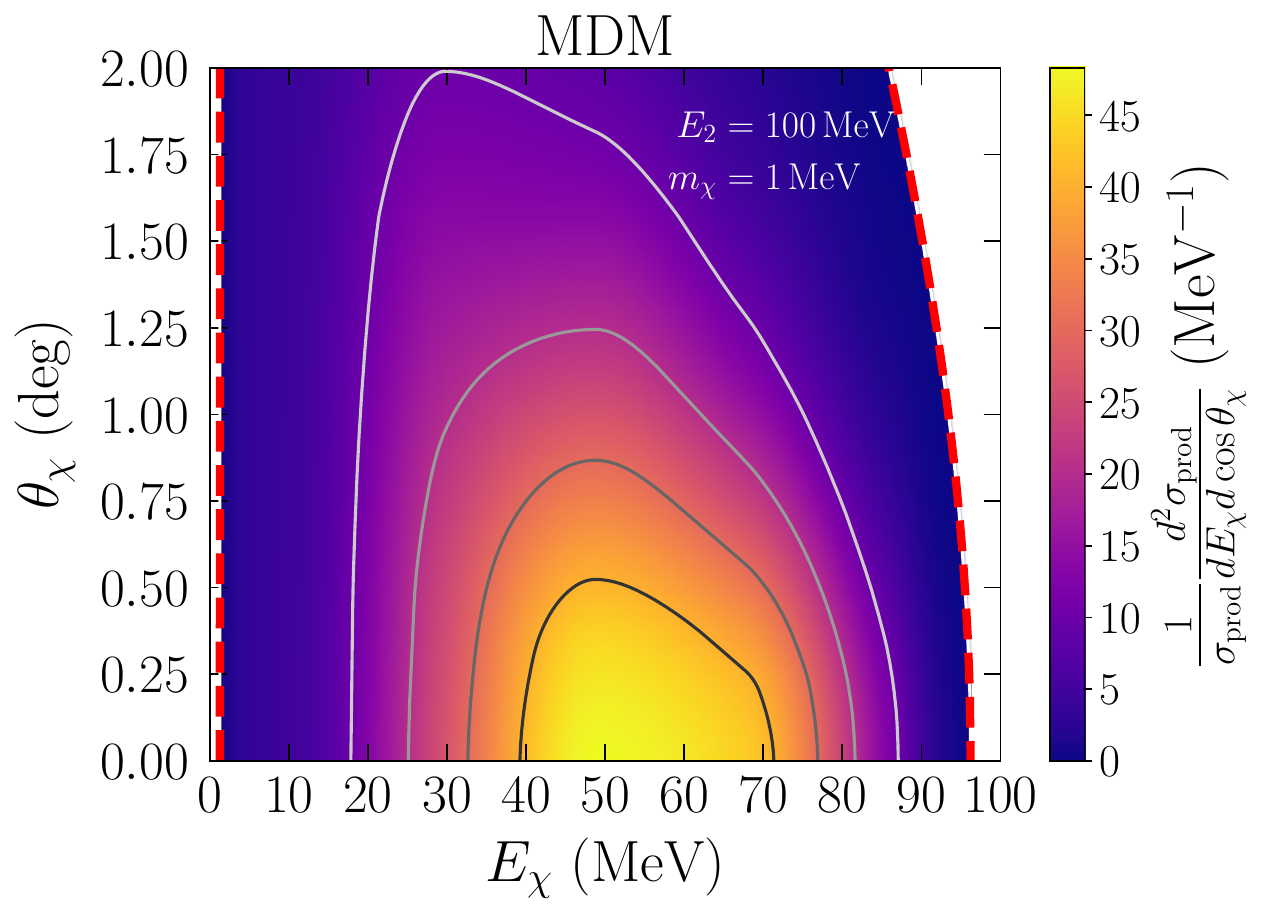}
	\end{subfigure}
	\begin{subfigure}[h]{0.495\textwidth}
		\centering
		\includegraphics[width=0.99\textwidth, height=!]{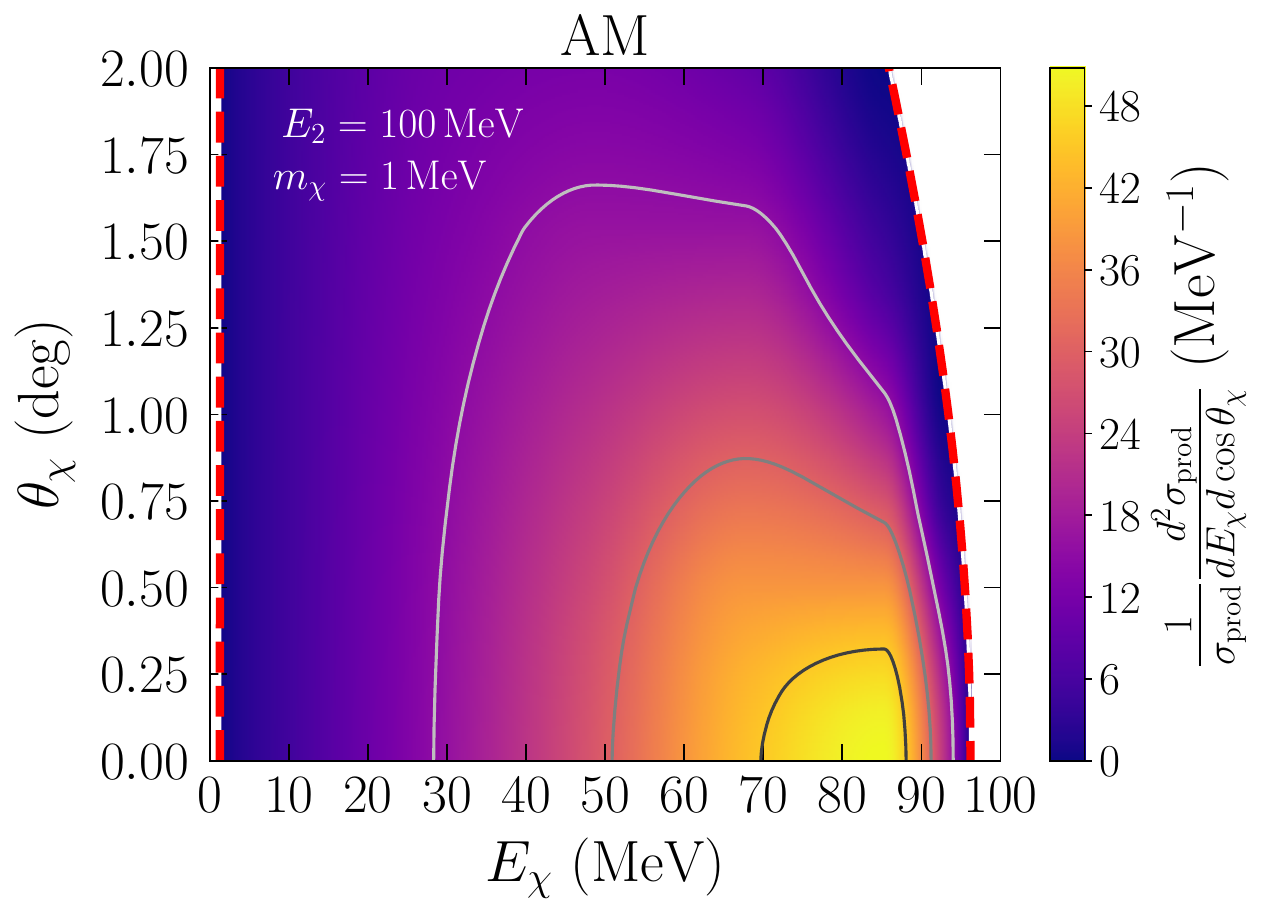}
	\end{subfigure}
	\caption{Heat maps showing the normalised differential production cross section of a particle of mass $m_{\chi}=1\,$MeV with respect to the energy $E_{\chi}$ of the produced particle and its angle $\theta_{\chi}$. The beam energy is set to $E_2=100$\,MeV. The \emph{left panel} shows the result for MDM and the \emph{right panel} for AM. The differential cross section is dominated by the angular region of $0^\circ$ to $2^\circ$, similar to the millicharged case. However, while the dominant energy region for MDM is also comparable to the millicharged case, for AM the region is narrower and shifted towards higher energies.}
	\label{fig:fm_geo}
\end{figure}

Figure~\ref{fig:fm_geo} shows the differential production cross section as a heat map with respect to the angle and energy of the $\chi$ particle with EM form factor interactions. The left panel of Fig.~\ref{fig:fm_geo} shows MDM, which represents a mass-dimension~5 operator as well as one without~$\gamma^5$. Compared to the millicharged case in Fig.~\ref{fig:milli.geo}, MDM peaks around slightly larger energies and angles. The right panel of Fig.~\ref{fig:fm_geo} depicts AM, representing a mass-dimension~6 operator with~$\gamma^5$. The higher dimension of the operator shifts the dominantly contributing region to higher energies, close towards the kinematic endpoint. The $\gamma^5$ of AM also leads to a narrowing of the energy region and a preference towards larger angles. The dominant angular region for the differential production cross section, however, remains around $<2^\circ$.

\bibliographystyle{JHEP}
\bibliography{ref}

\end{document}